\documentclass[preprint2]{aastex}

\usepackage{natbib}
\usepackage{graphicx}
\usepackage{txfonts}

\slugcomment{}

\shorttitle{ }
\shortauthors{Chabot et al.}

\usepackage{xcolor}

\begin{document}

\title{Reactions forming C$_{n=2,10}^{(0,+)}$, C$_{n=2,4}$H$^{(0,+)}$ and C$_{3}$H$_2^{(0,+)}$ in the gas phase:\\ semi empirical branching ratios.}

\author{M. Chabot\altaffilmark{1}, K. B\'eroff\altaffilmark{2}, P. Gratier\altaffilmark{3}, A. Jallat\altaffilmark{1}, V. Wakelam\altaffilmark{4,5} }
%\email{chabot@ipno.in2p3.fr}
\altaffiltext{1}{Intitut de Physique Nucl\'eaire d'Orsay,IN2P3-CNRS and Universit\'e Paris-Sud, 91406 Orsay cedex, France}
\altaffiltext{2}{Institut des Sciences Mol\'eculaires dOrsay, CNRS and Universit\'e Paris-Sud, 91405 Orsay cedex, France}
\altaffiltext{3}{Institut de Radioastronomie Millim\'etrique, 300 rue de la Piscine, 38406 Saint Martin d'H\`eres, France}
\altaffiltext{4}{Univ. Bordeaux, LAB, UMR 5804, F-33270, Floirac, France}
\altaffiltext{5}{CNRS, LAB, UMR 5804, F-33270, Floirac, France }

\begin{abstract}

The aim of this paper is to provide a new set of branching ratios for interstellar and planetary chemical networks based on a semi empirical model. We applied, instead of zero order theory (i.e. only the most exoergic decaying channel is considered), a statistical microcanonical model based on the construction of breakdown curves and using experimental high velocity collision branching ratios for their parametrization. We applied the model to ion-molecule, neutral-neutral, and ion-pair reactions implemented in the few popular  databases for astrochemistry such  as KIDA, OSU and UMIST.  We studied the reactions of carbon and hydrocarbon species with electrons, He$^+$, H$^+$, CH$^+$, CH, C, and C$^+$ leading to intermediate complexes of the type C$_{n=2,10}$, C$_{n=2,4}$H, C$_{3}$H$_2$, C$_{n=2,10}^+$, C$_{n=2,4}$H$^+$, or C$_{3}$H$_2^+$ . Comparison of predictions with measurements supports the validity of the model. Huge deviations with respect to database values are often obtained. Effects of the new branching ratios in time dependant chemistry for dark clouds and for photodissociation region chemistry with conditions similar to those found in the Horsehead Nebula are discussed.   
\end{abstract}


\section{Introduction}

	Carbon nuclei are produced by the triple alpha reaction \citep{1952ApJ...115..326S} inside stars during  Helium burning.   Dying stars in the asymptotic giant branch (AGB) and super nova explosions (SNe) II are the two main classes of stars that are responsible for the total amount of carbon into the galaxy today \citep{clayton2003handbook}.  In interstellar medium (ISM), carbon nuclei may be observed in atomic and ionic forms. They also form the backbone of a huge number of molecules observed in the gas phase, from the most stable diatomic ones such as CO \citep{2001ApJ...547..792D}  to the large polyatomic species such as PAH \citep{2008ARA&A..46..289T}.  Carbon is also contained in various condensed matter in the form of grains such as hydrogenated amorpheus carbon (HAC), carbon dusts, diamond, ... 
\cite[][and references therein]{2011EAS....46..381D}. Carbon is present in ices \citep{1991ApJ...381..181T} and in large carbon based molecular structures mixed with silicated material grains found in primitive carbonaceous meteorites on earth \citep{2007GeCoA..71.4380A}.
	 
	It is believed that carbon dusts are strongly produced by AGB stars and type II SNes \citep{2009MNRAS.396..918M}. In their release phases, ions, atoms, simple molecules and  more complex ones are also synthesized. From star ejecta three types of carbonaceous matter (i.e. atomic, molecular and dust) is injected in  the surrounding  ISM and will dynamically evolve.   The solid reservoir may be transformed partly or totally in molecular form and/or atomic form by shocks \citep{1990QJRAS..31..567J, 1994ApJ...431..321T}   or/and UV \citep{1997ApJ...489L.193S}  and/or particle collisions \citep{2011A&A...529A.146G}.  On the other hand, atoms and/or molecules could condensate to form  solids and ices on existing grains \citep{1999SSRv...90..219A}. Chemical reactions in the gas phase will modify the chemical composition of ISM. Reaction of atoms and/or molecules at the surface of (icy) grains  will play also  an important role in the chemical evolution of the carbon \citep{2006A&A...457..927G}.  
	
	In all these complex and dynamically coupled processes, gas - phase chemistry is certainly the easiest subject to tackle. Indeed, since it involves two body reactions, measurements and/or calculations are possible with few assumptions. In contrast, chemistry at the surface of grains and chemistry resulting from   desorption/erosion are speculative since the composition and morphology of grains are poorly known. Thanks to intense studies by molecular physicists and/or chemists, large databases with thousands of reaction rate coefficients such as KIDA \citep{2012ApJS..199...21W}, OSU \citep{1980ApJS...43....1P}, or UDfA \citep{2007A&A...466.1197W, 2013A&A...550A..36M}   exist for gas-phase ISM chemistry.  The  new KIDA database is a unique open tool in which to incorporate modern experiments and calculations performed in the field.  

	 Laboratory measurements on all the possible reactions is clearly a very hard task even if sensitivity analysis \citep{2006A&A...451..551W} may reduce the number of "key" reactions that are really meaningful. To address this problem  statistical theory is commonly used to predict branching ratios (BR) in the outgoing reaction channels of type A+B $\rightarrow$ C+D \citep{1978ApJ...222..508H, Liu2005173}.  In databases,  if no detailed measurement or calculation exists, zero order statistical predictions are used which is equivalent to put all the reaction in the most exoergic channel. 
	 
   In the present work, we propose a semi empirical statistical model to go beyond this zero order approximation. This new approach uses the mass spectrometry concept of breakdown curves \citep{JMS:JMS354} and a large set of experimental branching ratios (BR) measured in high velocity collisions (HVC) at the Orsay Tandem facility. It is presented in  section two. We apply this model to all reactions mediated by an intermediate complex (A + B $\rightarrow$ AB$^*$) such as C$_n^{(0,+)}$, C$_n$H$^{(0,+)}$, C$_3$H$_2^{(0,+)}$. With respect to previous work on the same systems \citep{2010A&A...524A..39C} the present model is able to introduce the proper energy deposit associated to a particular reaction. Then it allows to treat not only  photodissociation and electronic dissociative recombination (DR) as previously published \citep{2010A&A...524A..39C} but also chemical reactions.  In the third  section, the BRs predicted by the model are presented and compared with the BRs from KIDA\footnote{http://kida.obs.u-bordeaux1.fr/}, OSU 01-2007\footnote{https://www.physics.ohio-state.edu/$\sim$eric/research.html} and UdFA06\footnote{http://www.udfa.net/} databases for DR, ion-neutral, neutral - neutral, and anion-cation molecular reactions. In the fourth section, effects of the new  BRs on the gas- phase chemistry are studied in  the dark Taurus Molecular Cloud 1 (TMC1) conditions \citep{1981A&A....95..143T} and in the photon-dominated region (PDR) Horsehead Nebula conditions \citep{2005A&A...435..885P}.

\section{Semi empirical statistical model}
\subsection{Principle}

In a statistical microcanonical formulation of the fragmentation \citep{2004PhRvL..93f3401M}, the branching ratio of a decaying channel j may be written:

\begin{equation}
 BR_j =  \int_0^{+\infty} BDC_j(E) \times f(E) dE
\end{equation}

Where $f(E)$ is the normalized internal energy distribution of the parent that undergoes fragmentation and  $BDC_j(E)$ is the internal-energy dependent dissociation  probability, also called breakdown curve for channel $j$, verifying at each energy:

\begin{equation}
 \sum_j BDC_j(E) = 1
\end{equation}

\begin{figure}
\begin{center}
\begin{tabular}{c}
\includegraphics[width=1.0\linewidth]{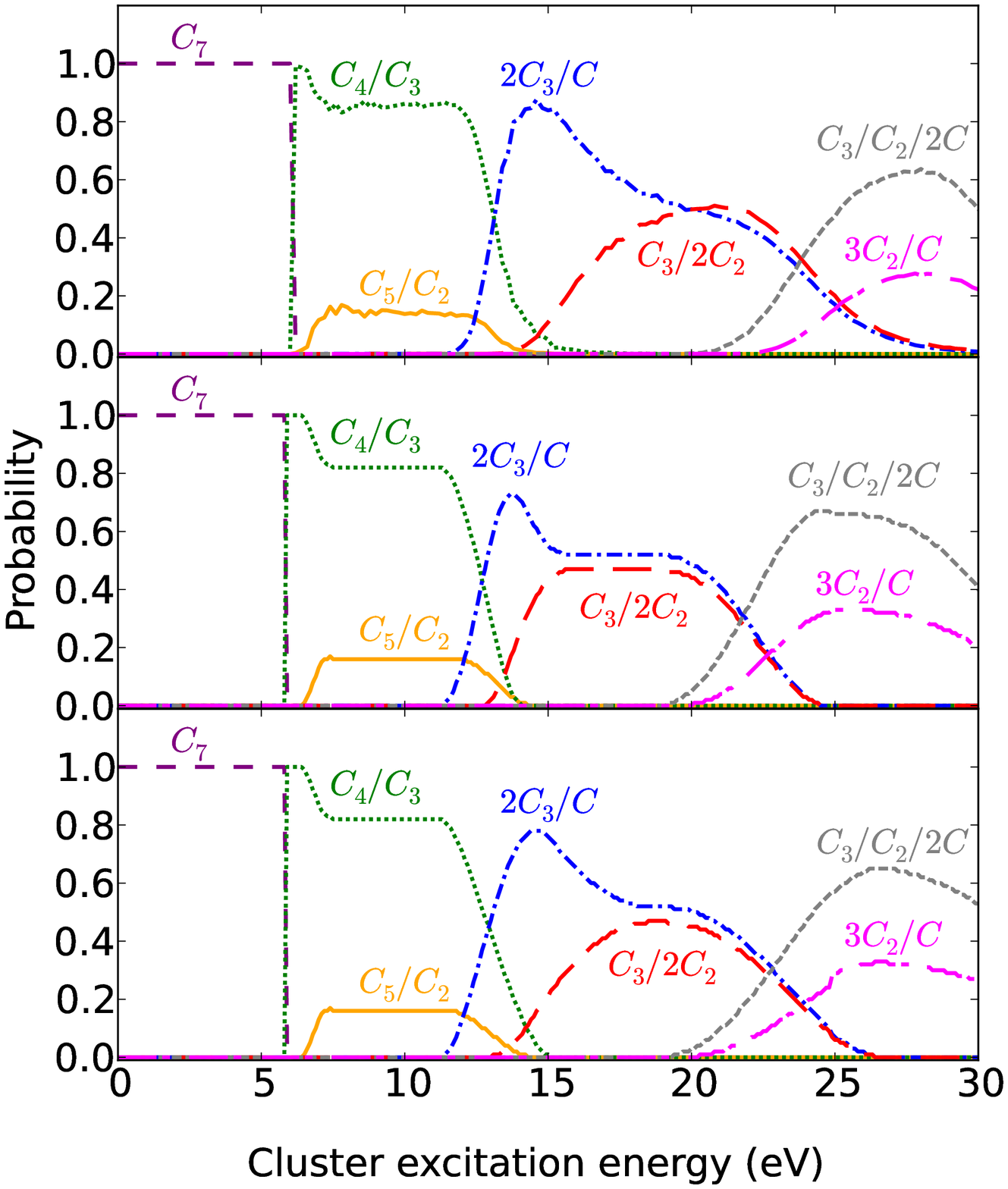} \\
\end{tabular}
\caption{Breakdown curves for the C$_7$ molecule. Upper panel: Theoretical microcanonical metropolis monte carlo (MMMC) calculation  \citep{2004PhRvL..93f3401M}. Middle panel: semi empirical model with E$_{sat}$ obtained with formulaes 4 and 5. Lower panel: semi empirical model with E$_{sat}$ adjusted inside the error bars (see \S 2.4.) to reproduce MMMC calculations.\label{BDCs_C7}}
\end{center}
\end{figure}

 The principle of our model is the following:  since we measure, in experiments, fragmentation branching ratios $BR_j$ for all channels, de-exciting a molecule whose internal energy distribution f(E) is known \citep{2008JChPh.128l4312T, 2010A&A...524A..39C}, we extract the  $BDC_j(E)$ for all j channels, by  equation (1) inversion. These $BDC_j(E)$ can then be used to predict BRs in a molecule possessing any internal energy, using equation (1), that means, for a variety of physical or chemical processes associated to very different energy deposits as will be shown.
The concept of breakdown curves, or breakdown diagram, is well known in mass spectrometry (see for instance \cite{JMS:JMS354}) and has been recently used for theoretical interpretation of fragmentation branching ratios of neutral \citep{2004PhRvL..93f3401M} or multi-charged \citep{2010PhRvL.104d3401C} carbon clusters. In figure~\ref{BDCs_C7}  an example of breakdown curves calculated within the microcanonical metropolis monte carlo (MMMC) method for the case of C$_7$ fragmentation  is shown \citep{2004PhRvL..93f3401M}. This work uses an extensive phase space and extract density of states on the basis of high precision level \textit{ab initio} calculations \citep{2005PhRvA..71c3202D}. In figure~\ref{BDCs_C7}, these MMMC BDCs are compared to BDCs derived with the present model by inversion of equation (1) and using a physical parametrization of $BDC_j(E)$ as detailed below. Both  approaches  give very similar results. 

\subsection{Construction of the breakdown curves}

\begin{figure}
\begin{center}
\begin{tabular}{c}
\includegraphics[width=1.0\linewidth]{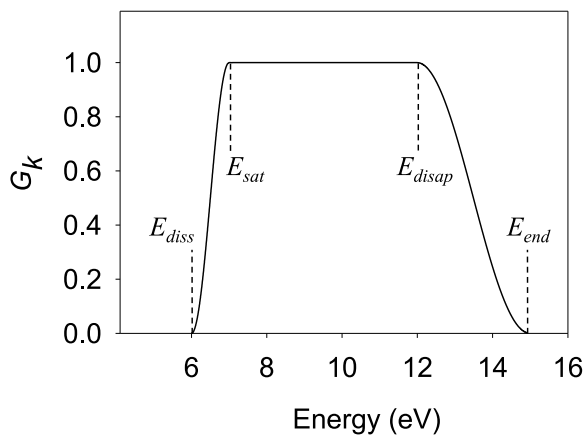} \\
\end{tabular}
\caption{Generic form of the G$_j$ function. \label{Gk} }
\end{center}
\end{figure}

The energy dependence of $BDC_j(E)$  is easily understandable. When the parent internal energy is below the energy needed for the channel j to occur, the probability is zero. When the energy is above the energy needed for \textit{additional} dissociation of the daughter or of one of the fragments, the probability of channel j is decreasing rapidly to zero. Between the two, the probability is either 1 if this dissociation channel is unique or less than one if there is competition with other dissociation channels. Accordingly, we expressed the $BDC_j(E)$  curve as follows:

\begin{equation}
 BDC_j(E) = \frac{a_j \times G_j(E)}{\sum_j{a_j \times G_j(E)}}
\end{equation}

where $G_j(E)$ has the generic form depicted in figure~\ref{Gk}, $a_j$  takes into account the possible competition of channel $j$ with other opened channels and the denominator ensures normalization (2). 

The physical interpretation of $G_j(E)$ is the following: it represents the probability of dissociation along channel $j$ when not in competition with another channels with the same number of fragments. The rise above E$_{dis}$ and the decrease above E$_{disap}$ of $G_j(E)$ (see figure 2) are due to the competition with channel having a different number of fragments. The rise and decrease of the $G_j(E)$ function are usually steep which can be easily explained.  Indeed, \cite{JMS:JMS354} has shown that the ratio between probabilities of competitive channels is equal, at infinite times, to the ratio between corresponding rate constants and it is well  known that rate constants (and their ratios) do vary  rapidly at the opening (or closing) of a channel \footnote{Expression of the rate constant for dimer and trimer evaporation from a cluster having internal energy below or above the opening of further dimer and trimer dissociation may be found for instance in \cite{2005PhRvA..71c3202D} (equations (31) and (35)).} whereas ratios tend to be energy independent elsewhere.

The $G_j(E)$ function depends apparently on four parameters: E$_{dis}$, E$_{sat}$, E$_{disap}$, and E$_{end}$.  In fact, the decrease of $G_j(E)$ is complementary to the increase of the daughter fragmentation channel, as mentioned before. It makes E$_{disap}$ to be equal to the lowest E$_{dis}$ of the daughter fragment and E$_{end}$ to be equal to the corresponding E$_{sat}$. We are then left with the determination of E$_{dis}$ and E$_{sat}$ for a particular dissociation channel. E$_{dis}$ corresponds to the minimum energy needed for this channel to occur, i.e, the dissociation (or formation) energy that is usually known from literature (see \S 2-3). The E$_{sat}$ value and the variation between E$_{dis}$ and E$_{sat}$ is more difficult to evaluate. The competition between channels with a different number of emitted fragments is explained by the different partitioning of the parent internal energy. Indeed, this energy may be used for production, motion and internal exitation of the fragments. We used MMMC breakdown curves as a guide to estimate E$_{sat}$ values. From this theoretical work \citep{2006IJMSp.252..126D} we could deduce that:

\begin{eqnarray}
E_{sat} \approx E_{dis} + 1.0 eV  &  if & N_F = 2
\end{eqnarray}
\begin{eqnarray}
E_{sat} \approx E_{dis} + (N_F - 2) \times 1.5 eV &  if & N_F > 2 
\end{eqnarray}

where $N_F$ is the number of emitted fragments in the considered channel. Dispersion of E$_{sat}$-E$_{dis}$ values from channel to channel was found of the order of 25\%.

On the other hand, the slope of the curve between E$_{dis}$ and E$_{sat}$ was modelled by the following function:

\begin{equation}
G_j = \frac{1}{2} \times (1.-\cos ( \pi \times \frac{E_{dis}-E}{E_{sat}-E_{dis}})) 
\end{equation}

Having constructed the $G_j(E)$, the $a_j$ scaling factors were extracted by minimization between measured and predicted BRs using equation (1)(2) and (3). By this manner semi empirical $BDC$s were obtained.

\subsection{Model inputs}

We used  dissociation energies for C$_n$ and $C_n^+$  taken  from \cite{2006IJMSp.252..126D} and \cite{2006BrJPh..36..529D}. They have applied the density functional theory (DFT) with the B3LYP functional for exchange and correlation. For C$_n$H we used dissociation energies of \cite{2003JChPh.119.7705P} and for C$_3$H$_2$ those of \cite{ISI:000074301600017}. For dissociation energies of C$_n$H$^+$ and C$_3$H$_2^+$ species we combined those last neutral dissociation energies with the ionisation potentials from \cite{ISI:A1992GY63100014}.
  
The set of experimental BRs that we used to adjust the scaling factors $a_j$ in the model  were obtained with HVC experiments. Part of them has already been  published in \cite{2008JChPh.128l4312T} (Tables II to V: $C_nH$; Tables VII to X: $C_nH^+$ n=1, 4 ) and in \cite{2006JPhB...39.2593C} (Tables 1 to 10 : C$_n, $ n=2, 10).  All other unpublished experimental BRs concerning C$_n^+$,C$_3$H$_2$ and C$_3$H$_2^+$ are given in  supplementary material available online. Details on the measurements for those last species may be found in \cite{2006JPhB...39.2593C} and in \cite{2008JChPh.128l4312T}.

Determination of the internal energy distribution resulting from the high velocity collision, $f(E)$,  was done using multiplicity (i.e probability distribution in numbers of fragments) as detailed in \cite{2008JChPh.128l4312T} and \cite{2006IJMSp.252..126D}. This is an improvement over the so called thermometer-method \citep{Wysocki1987181} currently used in mass spectrometry.

\subsection{Model confidence and error generation}

For chemical reactions, the model does not take into account any rule other  than the energy conservation. We checked (see appendix A) whether symmetry considerations or spin and angular momentum conservation rules could lead to forbidden transitions of energetically allowed chemical reactions.  In Tables of section 3, we indicated the channels where problems may arise. The characteristics of the ground states of reactants and products are given in Table~\ref{sym}.

\begin{figure}
\begin{center}
\begin{tabular}{c}
\includegraphics[width=1.0\linewidth]{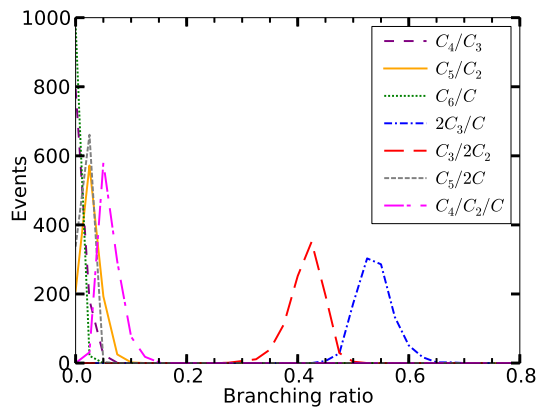} \\
\end{tabular}
\caption{Example of BR distributions obtained when introducing the error on  E$_{sat}$ - E$_{dis}$ (see $\S$ 2.4). Internal energy of the C$_7$ molecule was taken equal to 13.87 eV in the model.\label{BR_MC}}
\end{center}
\end{figure}

Measured BRs used in equation (1) (left-hand side) have their own error bars that are almost identically retrieved in BR’s predictions. These errors are typically, in absolute, less than 0.03 for C$_n$ species, less than 0.05 for C$_n$H species and close to 0.10 for a few channels in the case of C$_3$H$_2$ measurements. On the other hand, the sensitivity of the model to the used dissociation energies and E$_{sat}$ values was checked. For dissociation energies, we run calculations with either DFT-B3LYP or CCSDT energies \citep{PhDST}, those typically differ by 1 ev in absolute and by less than 0.1 eV in relative. Changes on predicted BR’s have been found to be less than 0.05 in absolute in the worse cases.  For E$_{sat}$,  we solved the inversion of equation (1) by introducing a gaussian distribution of (E$_{sat}$-E$_{dis}$) values with a standard deviation taken equal to 25 \% of the peak value.  Figure~\ref{BR_MC}  presents an example of BR distributions that were obtained for C$_7$ fragmentation and from which error bars were derived. In Tables of section 3 we give these errors for all channels.  In the end, the sum of all sources of errors makes the model more a qualitative tool designed to correct unrealistic predictions sometimes present in astrochemical databases (see \S 3) than a precise quantitative ones.

\section{Branching ratios model predictions}

\subsection{Dissociative recombination}

Branching ratios for dissociative recombination (DR)  may be predicted with the model assuming that the internal energy of the neutral intermediate complex is equal to its first ionization potential (IP).  Calculated BRs with this assumption are reported for C$_n^+$, C$_n$H$^+$ and C$_3$H$_2^+$ species in Tables~\ref{DR1},~\ref{DR2} and ~\ref{DR3} respectively. In the same tables, BRs found in the most commonly used databases in astrochemistry are reported. We also report the exothermicity in all the outgoing channels. 

For C$_n^+$ and C$_3$H$^+$, since IPs are lower than the lowest energy of three fragment dissociation,  only two fragment dissociation channels have to be considered. The model results are very close to previous published BRs \citep{2010A&A...524A..39C}, that have been included in the KIDA database. These values differ strongly from predictions of other databases, which clearly underestimate channels with the magical C$_3$ fragment. For C$_4^+$, all databases report the measurements from \citet{heber:022712} and the model is in good agreement with those values.

For the DR of  C$_2$H$^+$ and C$_4$H$^+$ molecules, due to high values of IPs, three fragment channels are open. The model predictions for C$_2$H$^+$ are close to the experimental results of \citet{2004PCCP....6..949E}. The model confirms the observed but unquantified three fragment channel in experiments by \cite{Angelova2004195} for C$_4$H$^+$.   Note that in all cases the errors on the predicted BRs for the three fragment channels are quite large. It is because the IP lies just above the opening of the three fragment channels.
															      
The C$_3$H$_2$ molecule has two isomers in astrochemistry databases. One is cyclic and the other is linear. Since these two forms differ in formation energy by only  0.1 eV \citep{PhDTT}, the dissociation energies used in the model were considered to be identical. Note that the  C$_3$/H$_2$ channel has a formation barrier \citep{2008P&SS...56.1658L}  and we used it as the dissociation energy in the model. In contrast with  formation energies,  IPs are quite different between linear and cyclic forms. As can be seen in Table~\ref{DR3}, DR- BR are not identical for the two isomers. For the linear form, three fragment channels are predicted to be more populated. Note that the errors on the model are quite large in both cases because internal energies are close to the opening energies of the three fragment channels.

\subsection{Ion-molecule reactions}

\subsubsection{Charge exchange reaction with He$^+$ }
	   
Exothermic charge exchange reaction between molecules and He$^+$ occurs with large reaction rates, as  some electronic states in the inner valence shells of the molecule are energetically close to the Helium IP \citep{fisher:2296}. This makes the charge transfer resonant. Since in the output channel, He atom  is likely to be produced  with only little kinetic energy, the internal energy of the neutral molecule is assumed to be close to $\Delta$IP = IP(He) - IP(molecule). Tables~\ref{He1},~\ref{He2}, and ~\ref{He3} give the BRs calculated by the model for  C$_n$, C$_n$H and C$_3$H$_2$ charge exchange with He$^+$. Internal energies reported in the same tables were obtained using the IPs reported in Tables~\ref{DR1},~\ref{DR2}, and ~\ref{DR3} with IP(He)=24.58 eV. 

In all the cases, because the IP of helium is high, the internal energy of the charged molecule is also high. As a consequence, fragmentation into three fragments operates, although neglected in all databases. For C$_4$, it is small, but it increases with the size and reaches half of the probability for C$_{10}$.  Nevertheless $\Delta$IPs are close to the dissociation energies of three fragment channels, therefore errors are quite important. For example for C$_6$ species,  using CCSDT calculations for dissociation energies  instead of DFT-B3LYP \citep{PhDST}, we find that almost all the dissociation goes into three fragment channels. Here again, all reported BRs in databases miss the importance of the channels with a C$_3$ emission. For the C$_n$H, the three fragment channel contribution is also increasing with the size and reaches almost 100\% for C$_4$H. Differences between the model predictions and the values reported in databases are huge. It is because in databases: (i) only two fragment channels are considered and (ii) the H emission is strongly overestimated. 

Since in databases cyclic and linear isomers of C$_3$H are separated, we report BRs using the same dissociation energies but a different IP for the linear and cyclic species. As a consequence, the linear species produces more three fragment channels than the cyclic one. For C$_3$H$_2$, all the probability goes into the three fragment channels.
	     
\subsubsection{Ion neutral bimolecular reactions}
 
For bimolecular reactions between neutral and ionic species, the internal energy of the intermediate complex results from the association. Therefore it is calculated  by the dissociation energy of the reverse pathway. For instance for C$_n$ + C$^+$ reaction, the internal energy of the complex (E$_{ass}$) is equal to the dissociation energy of C$_{n+1}^+$ into C$_n$ + C$^+$.  Tables~\ref{ione1} to~\ref{ione6} give the BR calculated by the model for intermediate complexes  of type C$_n^+$, C$_n$H$^+$ and C$_{3}$H$_2^+$. 

For the reactions between C$_n$ species and C$^+$ displayed in Table~\ref{ione1}, the radiative deexcitation channel is always the only output channel in the reported databases. This is true for C$_3$ + C$^+$ since in that case internal energy in the intermediate complex C$_4^+$ is below all the energies of dissociation. 
For all other C$_n$ + C$^+$ reactions, the internal energy in the intermediate complex is above dissociation energy and those dissociation channels must be considered.  The lifetime of an intermediate complex is related to the energy difference between the internal energy and the dissociation energy. If the energy is close to the threshold, the complex will survive and will emit photons. While, if the energy is well above threshold it undergoes fragmentation. Using Weisskopf calculations \citep{2005PhRvA..71c3202D} we showed that, for C$_n^+$, fragmentation occurs faster  than  $10\mu$s/$100\mu$s for a few tenths of eV above threshold \citep{PhDST}. Whereas photon emission is on the millisecond time scale (Parneix P., private communication). Then, for all reactions presented in Table~\ref{ione1} fragmentation dominates, which is in  strong disagreement with databases assumptions. 

For  the C$_2$H$_2$ + C$^+$ reaction (Table~\ref{ione2}), the model agrees with experiments of \cite{doi:10.1021/j100402a038}.  

Concerning the reactions with H$^+$ (Table~\ref{ione3}), two channels have a very low exothermicity ( $ < $ 0.2 eV). For those two,  model results are very uncertain. It is noticeable that for C$_4$ + H$^+$ reaction, charge transfer is not the dominant channel. 

Looking at Table~\ref{ione4}, it appears that many of the reactions C$_n$ + C$_2^+$ are not implemented in databases. Pseudo time dependant chemical models for dense clouds \citep{1986MNRAS.222..689H} predict similar abundances for C$_2$ and C$_3$. Similarly, C$_3^+$  and C$_4^+$ would have abundances close to C$_2^+$. Thus reactions between C, C$_2$ and C$_3$ with C$_3^+$ or C$_4^+$ may be important to include into the models. The same remark holds for the reactions of C$_2$H, C$_3$H  with C$_2^+$, C$_3^+$, C$_4^+$, which are not implemented either.

For the reaction of C$_2$H + CH$^+$ (Table~\ref{ione5}), H$_2$ production predicted by the model is small due to the energy barrier as mentioned previously.  

\subsection{Neutral-neutral reactions}

 For neutral - neutral  reactions, as in the case of ion-neutral reactions,  internal energy of the intermediate complex (E$_{ass}$) is given by the dissociation energy of the reverse pathway. Tables~\ref{nene1} and~\ref{nene2} give the model BRs for the C$_n$H$_m$ intermediate complexes together with database BRs and exothermicities. The agreement between model predictions and reported values from KIDA and OSU databases is  pretty good due to a recent update based on statistical calculations \citep{2009A&A...495..513W}. The C + C$_2$H$_2$ reaction (Table~\ref{nene1}) has been extensively studied both experimentally and theoretically \cite[][and references therein]{2009ApJ...703.1179C}. The  BR of the C$_3$/H$_2$ channel  has been found to be strongly related to the collisional energy. At 0K (values reported in table 13), the BR is large (0.73) and it decreases down to 0.2 above 50 K. The model fails to reproduce the values at 0K and the temperature dependence. Note that this reaction is complicated, involving both a barrier on the outgoing C$_3$/H$_2$ channel and a intersystem crossing between triplet and singlet state of the intermediate complex \citep{doi:10.1021/jp0776208}.   

\subsection{Ion-ion reactions}

For ion pair recombinations of type A$^+$ + B$^-$ $ \rightarrow$ (AB)$^*$ $ \rightarrow$ C + D,  internal energy of the intermediate complex that undergoes fragmentation is given by :
E$_{dis}^{A/B}$ + IP$^A$ - EA$^B$, with E$_{diss}^{A/B}$: the energy of dissociation of AB into A/B, IP$^A$ : the ionization potential of the fragment A and  EA$^B$ : the electron affinity of the fragment B.  Table ~\ref{ioio1} gives the model BRs for C$_n$H$_m$ intermediate complexes.  These reactions are so far not implemented into the UDfA database. Internal energy of all  intermediate complexes are very high and three fragment channels are dominating for all sizes. Values reported in databases have clearly to be re-evaluated.

\section{Application to ISM chemistry}

Effects of the new branching ratios were studied using chemical models for two different environments: dark clouds and photon-dominated regions.

\subsection{Dark clouds}

To compute the chemical composition of a dark cloud, we used the Nautilus gas-grain model \citep{2009A&A...493L..49H}. This code takes into account the gas-phase chemistry, the sticking of gas-phase species to the surface, the evaporation of species from the surface and the surface reactions using the rate equation approximation. Details on these processes are given in \citet{2010A&A...522A..42S}. Species are initially in the atomic form, except hydrogen which is molecular. The elemental abundances are those of \citet{2012PNAS..10910233D}. The cloud temperature (gas and dust) is 10~K, the H density is $2\times 10^4$~cm$^{-3}$, the cosmic-ray ionization rate is $1.3\times 10^{-17}$~s$^{-1}$ and the visual extinction is 30. The gas-phase reactions are based on the kida.uva.2011 network \citep{2012ApJS..199...21W} while the surface network is the same as \citet{2007A&A...467.1103G}. 
 In the gas-phase network, we have modified the branching ratios according to the new values listed in Tables~\ref{DR1} to ~\ref{ioio1}. BRs for which the KIDA values have been underlined correspond to experimental results that were kept in the analysis i.e. not replaced by the model values.

\begin{figure*}
\epsscale{2}
\plotone{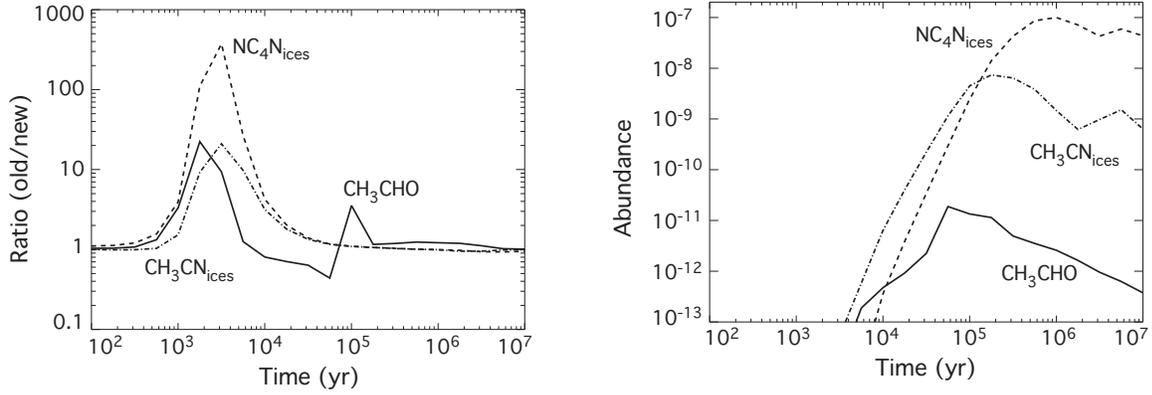}
\caption{Left: ratio between the species abundances computed with the kida.uva.2011 network (old) and the new  branching ratios (new)  . Right:  predicted abundances using the new branching ratios as a function of time are presented for selected species. Calculations are for a dark cloud. \label{figure_cloud}}
\end{figure*}

Effects of the new BRs are illustrated in Fig.~\ref{figure_cloud} for three selected species. These molecules have been chosen because they present the largest differences in the abundances computed with the two networks.  The effect on the chemistry of dark clouds is mainly seen at early times i.e. before $2\times 10^4$~yr. It is because later on, hydrocarbon chemistry is occuring at the surface of grains and through negative species reactions.  The early formation of carbon chains, such as C$_n$ and HC$_n$N, are delayed with the new BRs.  As a consequence, for few species produced at the end of reaction chains with hydrocarbon, large delays  are obtained and peculiar old/new ratio shapes obtained. It is for instance the case of the CH$_3$CHO molecule populated from the C$_2$H$_5$ precursor through C$_2$H$_5$ + O $\rightarrow$ H + CH$_3$CHO chemical reaction (see left of Fig.~\ref{figure_cloud}). 

\subsection{Photon-dominated regions}

For photon-dominated regions modeling, we used the Meudon PDR code \citep[http//pdr.obspm.fr, ][]{2006ApJS..164..506L}. The Meudon PDR code consistently solves the radiative transfer from far UV to sub-millimeter, chemistry, and thermal balance in a 1D plan-parallel and stationary slab of dust and gas. The density and thermal structures where determined in a consistent way using a constant $4\times 10^{6} \rm~K~cm^{-3}$ value for pressure \citep{2005A&A...437..177H}. The intensity of the incident UV flux was set to 60 times the ISRF (in Draine units \citep{1978ApJS...36..595D}) \citep{2005A&A...437..177H}, and the cosmic ray flux used was $5\times 10^{-17}$ per H and per second  \citep{2009A&A...498..771G}. The initial abundances were those of \cite{2006A&A...456..565G}. These values correspond to the Horsehead Nebula. This object is used for a benchmark calculation since it is well known that within it, the PAH or/and carbon grain reservoir are believed to be the main production source of hydrocarbons \citep{2005A&A...435..885P, 2012A&A...548A..68P}.

The gas-phase chemistry is identical to the kida.uva. 2011 network and, as for dense clouds, the new branching ratios listed in Tables~\ref{DR1} to ~\ref{ioio1} have been introduced.

\begin{figure*}
\includegraphics[width=.49\textwidth]{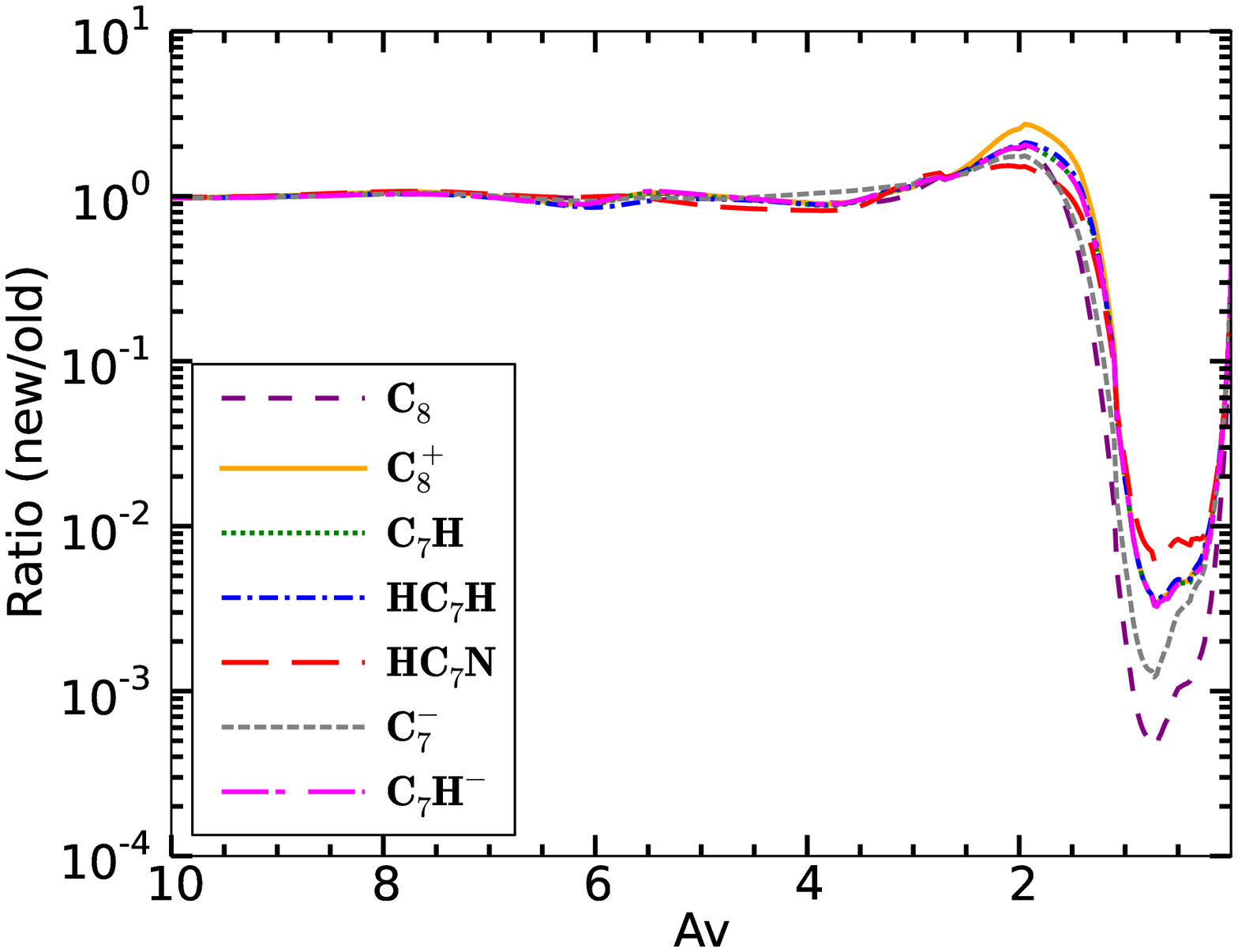}
\includegraphics[width=.49\textwidth]{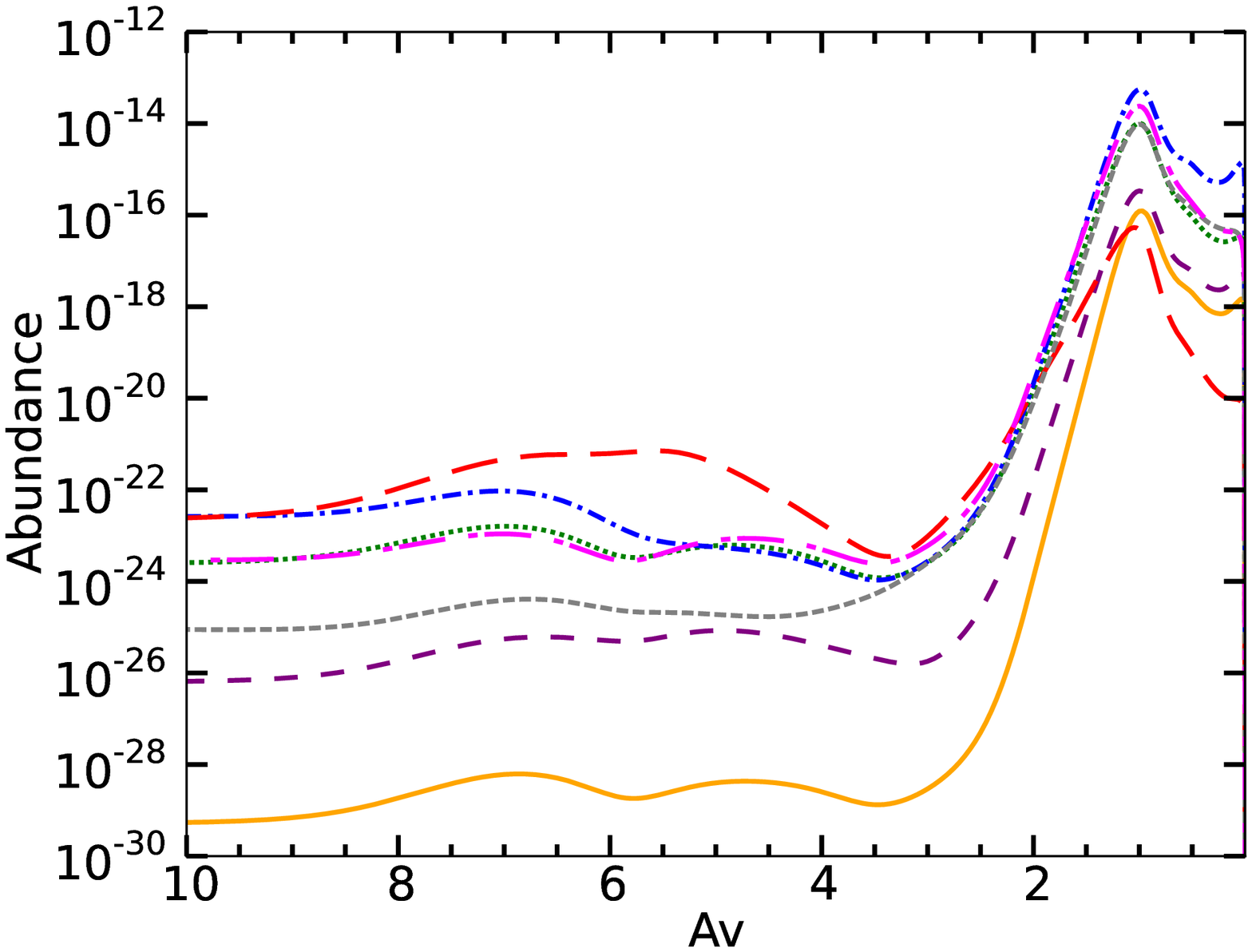} 
\caption{Left: ratio between the species abundances computed with  the new  branching ratios (new) and the kida.uva.2011 network (old). Right:  predicted abundances using the new branching ratios as a function of visual attenuation (Av).Calculations are for a PDR region with conditions similar to those found in the Horsehead Nebula. \label{figure_pdr1}}
\end{figure*}

\begin{figure*}
\includegraphics[width=.49\textwidth]{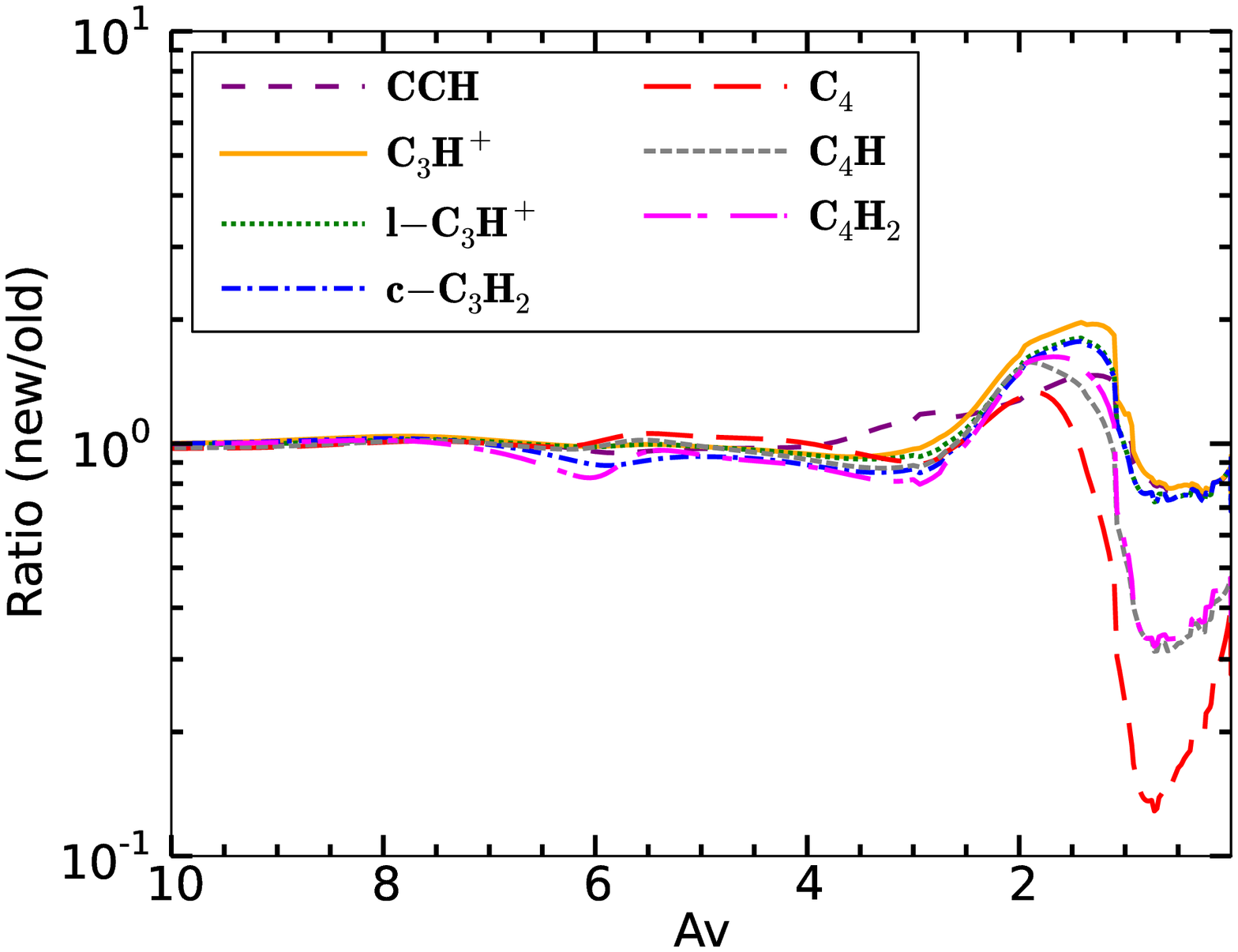}
\includegraphics[width=.49\textwidth]{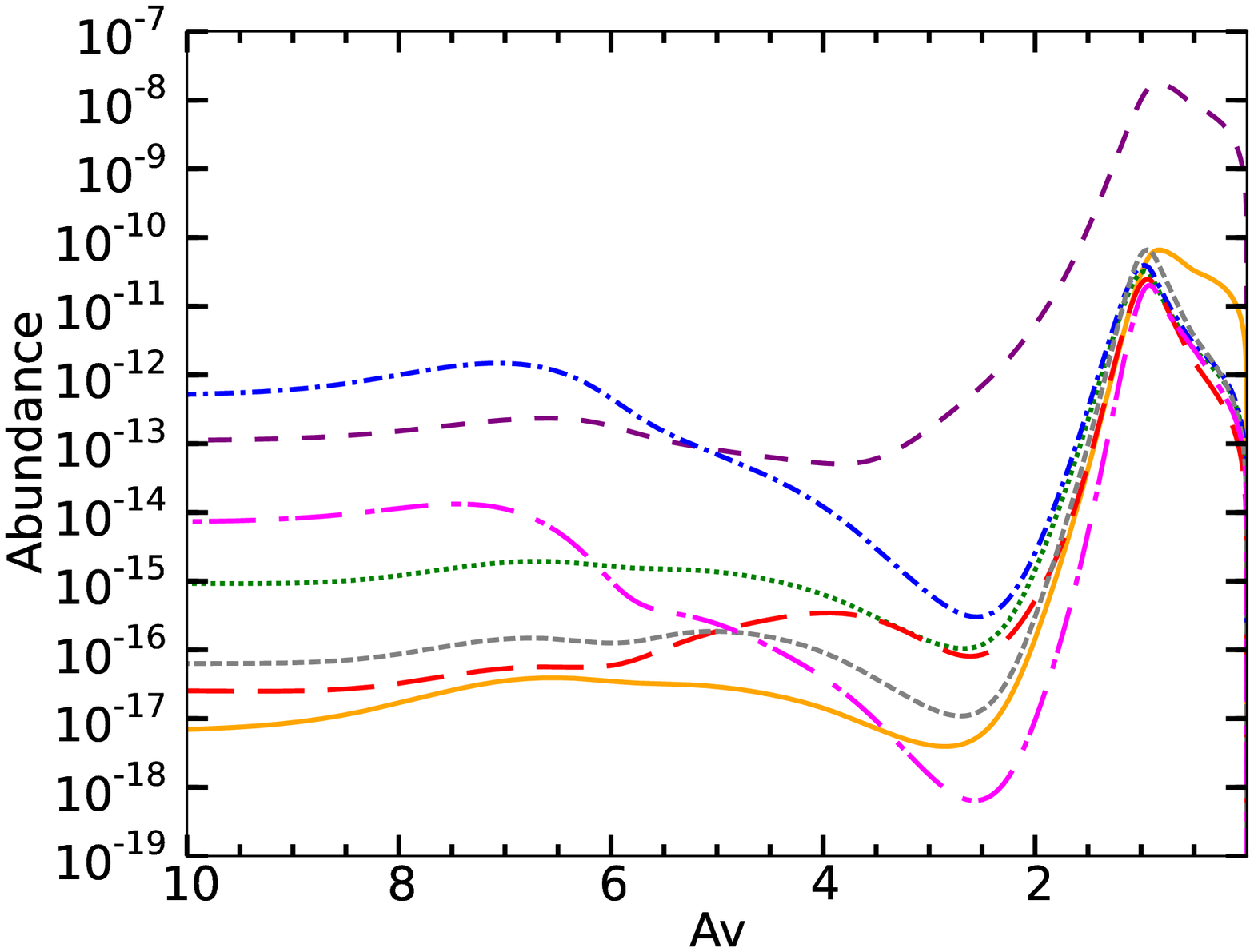} 
\caption{Left: ratio between the species abundances computed with  the new  branching ratios (new) and the kida.uva.2011 network (old). Right: predicted abundances using the new branching ratios as a function of visual attenuation (Av).Calculations are for a PDR region with conditions similar to those found in the Horsehead Nebula. \label{figure_pdr2}}
\end{figure*}

\begin{figure*}
\includegraphics[width=.49\textwidth]{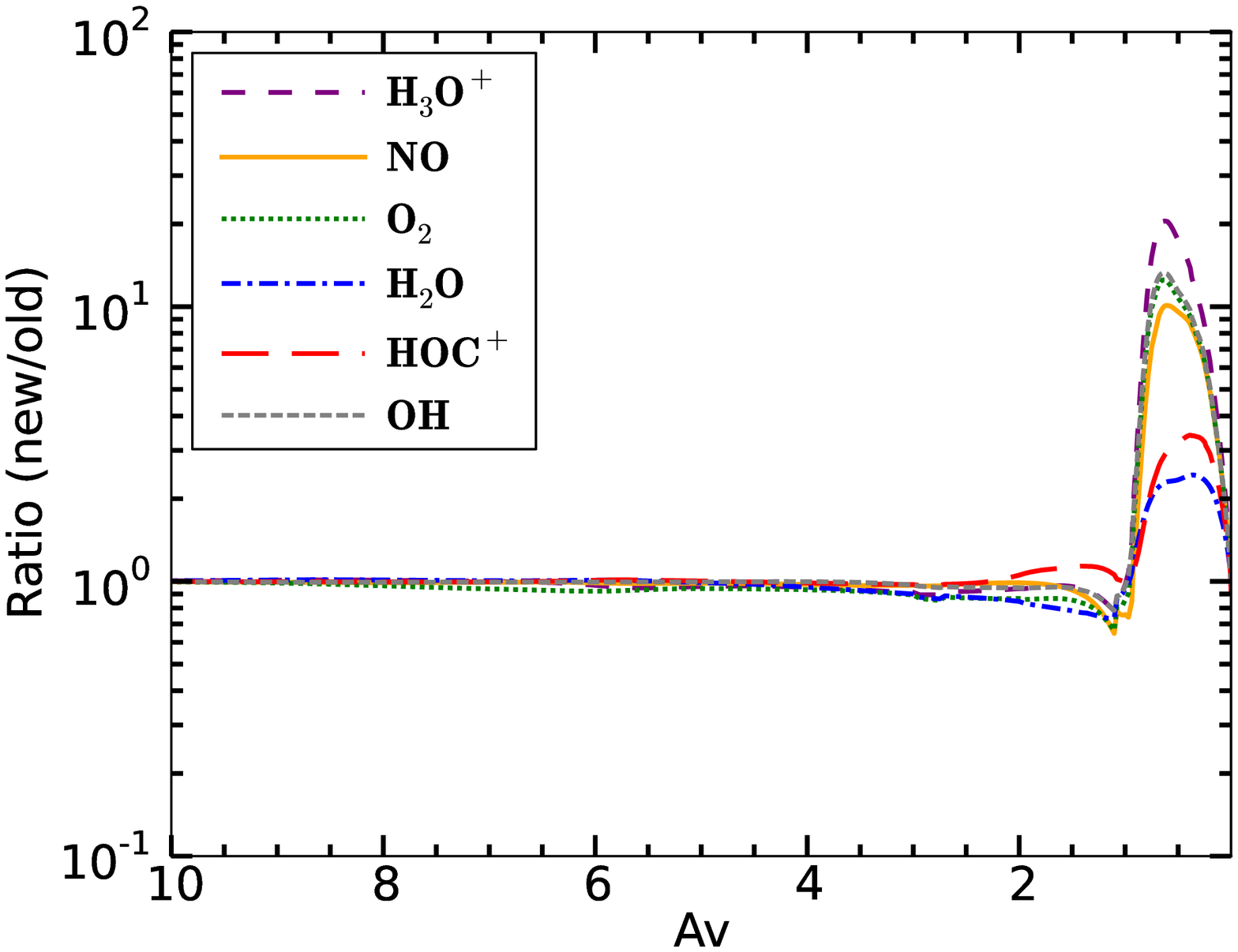}
\includegraphics[width=.49\textwidth]{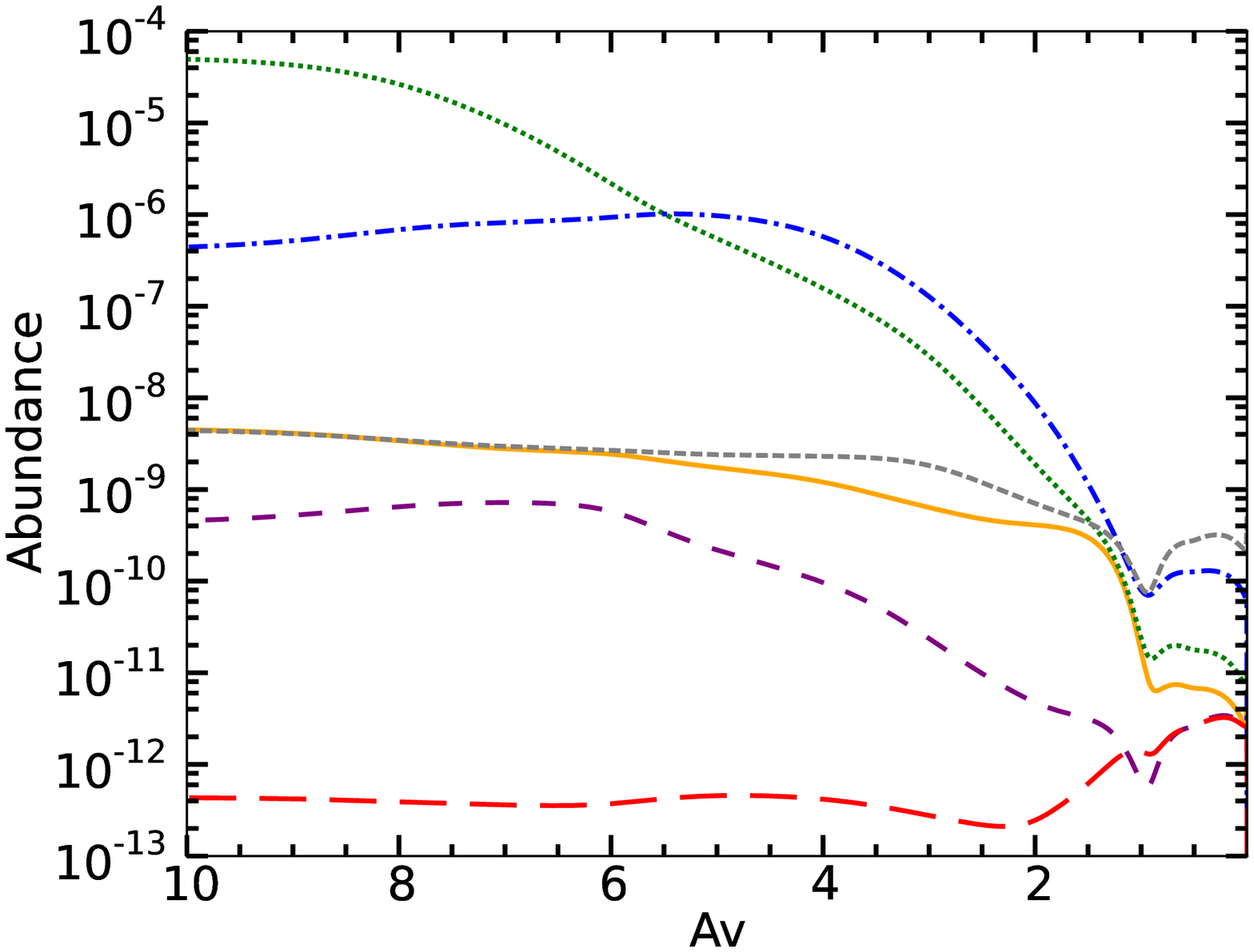} 
\caption{Left: ratio between the species abundances computed with  the new  branching ratios (new) and the kida.uva.2011 network (old). Right:  predicted abundances using the new branching ratios as a function of visual attenuation (Av).Calculations are for a PDR region with conditions similar to those found in the Horsehead Nebul. \label{figure_pdr3}}
\end{figure*}

Figures~\ref{figure_pdr1},~\ref{figure_pdr2}, and ~\ref{figure_pdr3} show, as a function of the visual extinction Av, the ratio of abundances computed with the new branching ratios over the ones computed with the old branching ratios for a selection of species.  The figure~\ref{figure_pdr1} focuses on some of the large hydrocarbon molecules.  Inside the cloud (Av $>$ 3) effects are very small. It is due to the fact that the large anion- neutral reactions, not modified in the present work, are the main pathways for the growth of large carbon chains \citep{2007ApJ...662L..87M}.  On the contrary, at the edge of the PDR  a large effect is observed and species abundances are strongly decreased by the new BRs. It is because in the old network C$_n$ species are locked in a loop that conserves the mass, this is : 

 C$_n$ + e$^-$  $\rightarrow$   C$_n^-$  +  $h\nu$
 
 C$_n^-$  +  $h\nu$ $\rightarrow$    C$_n$ + e$^-$
  
 C$_n^-$  +  C$^+$   $\rightarrow$  C$_n$  + C. 

With the new BRs (see Table~\ref{ioio1}) the rates of reactions  C$_n^-$  +  C$^+$   $\rightarrow$  C$_n$  + C  are put to zero and C$_n$s are fragmented. Therefore C$_n$s abundances decrease.  

Figure~\ref{figure_pdr2} presents the effect of the new branching ratios for small hydrocarbon molecules  observed in the real Horsehead Nebula PDR \citep{2004A&A...417..135T, 2005A&A...435..885P, 2012A&A...548A..68P}. These species are moderately affected by the new branching ratios. Ion - neutral reactions dominate the chemistry of these species together with dissociative recombination.  In the Tables~\ref{ione2} to ~\ref{ione5}, for small size hydrocarbons, the old and the new BRs have only few differences. 
 
Figure~\ref{figure_pdr3} presents the effect of the new branching ratios for the other most affected small species. It is noticeable that all contain oxygen.  They all derive from the production of O$^+$ resulting  from charge exchange between O and H$^+$. With the old network  H$^+$ is produced by cosmic rays and their induced photons while with the new network  CH + C$^+$ $\rightarrow$  C$_2$ + H$^+$  is a new pathway.  As a result H$^+$ abundance is enhanced and consequently the O$^+$ chemistry.  The error on branching ratio of this particular reaction is quite large and therefore deserves futher study. 

\section{Conclusions}

We introduced a statistical model to calculate semi empirical branching ratios based on high velocity collision experiments. We applied the model to many reactions leading to an intermediate excited complex of the type of C$_n^{+0}$, C$_n$H$^{+/0}$, and C$_3$H$_2^{+/0}$ that undergoes dissociation. We compared these new branching ratios to the ones from the most popular astrochemical databases. The new semi empirical branching ratios agree quite well with experimental BRs reported in databases with the notable exception of the C + C$_2$H$_2$ reaction at very low temperature where the model clearly failed. Since most of the BRs in the databases have been obtained with a statistical zero order hypothesis (i.e. all the reaction goes in the most exoergic channel), we propose to replace them by the new semi-empirical BRs which are by far more realistic.  We observed the effect of these new branching ratios in a time dependant chemical model under the physical condition of the dark cold cloud “TMC1”. No modification was found in the calculated abundances in the range of the 10$^5$ yr and more.  On the contrary  PDR calculations in the condition of the Horsehead Nebula for low Av extinction ( Av $<$ 2) with the same new branching ratios exhibit notable differences. It concerns  hydrocarbon synthesis but also species initiated by reactions with the H$^+$ ion such as oxygen based molecules. The new network will be available on the KIDA web site\footnote{http://kida.obs.u-bordeaux1.fr/}.

\acknowledgments

      All persons having worked in AGAT collaboration over the years are indebted for their valuable help in measurements. P. Pernot, P. Parneix and S. Diaz-Tendero are thanked for fruitful discussions. M. Eller is thanked for his careful reading of the manuscript. CNRS-IN2P3, CNRS-INSU, University of Paris-Sud have been the contributing partners. VW's research is funded by the French national program PCMI and the Observatoire Aquitain des Sciences de l'Univers. P. Gratier is funded by the grant ANR-09-BLAN-0231-01 from the French {\it Agence Nationale de la Recherche} as part of the SCHISM project (http://schism.ens.fr/).    

\appendix

\section{Appendix A. Experimental high velocity collision branching ratios}

The measured BRs for C$_n^+$, C$_3$H$_2$, and C$_3$H$_2^+$  are  displayed in Tables~\ref{C4p} to ~\ref{C10p}, Table~\ref{C3H2} and Table~\ref{C3H2p} respectively. All details concerning the experiments may be found in \cite{2006JPhB...39.2593C} and \cite{2008JChPh.128l4312T}.
  
\section{Appendix B. Correlation rules in chemical reactions}
 
We examine in this appendix whether chemical reactions between reactants X and Y in their electronic ground states leading to products Z and T in their electronic ground states are possible on the basis of correlation rules. These correlation rules, derived on the assumption of an adiabatic change of internuclear distances and in the frame of the Russell-Saunders coupling \citep{herzberg1939molecular}, have three origins:
\begin{enumerate}
\item [i)] Spin conservation

From the spin conservation and spin addition rules there must exist, for the reaction to proceed,  one value of the total spin S verifying both  conditions:  

$ \left| S(X)-S(Y) \right| \leq  S \leq S(X)+S(Y) $ 

$ \left| S(Z)-S(T) \right| \leq  S  \leq S(Z)+S(T)$

When this rule is not satisfied the reaction is spin-forbidden and preceded, in Tables~\ref{ione1} to ~\ref{ioio1}, by the sign $\dag$. The reaction may nevertheless occur through intersystem crossing.  

\item [ii)] Angular momentum conservation (linear molecules)

For linear molecules, similar rules hold for the projection, along the internuclear axis, of the angular momentum. For the reaction to proceed there must exist one projection of the total angular momentum M$_l$ such as: 

$  M_l=\left| M_l(X)+M_l(Y)\right|=\left| M_l(Z)+M_l(T)\right| $

Where M$_l$(X) is taking values between $-l_X$ to $+l_X$ if $X$ is an atom in a state of angular momentum l$_X$ and is equal to  $\pm\Lambda$ if X is a molecule in a state characterized by the $\Lambda$ quantum number ($ \Lambda = 0,1,2...$ for $ \Sigma, \Pi, \Delta...$ states) 
When this rule is not satisfied the reaction is preceded, in Tables~\ref{ione1} to ~\ref{ioio1}, by the sign $\S$.

\item [iii)] Symmetry considerations

Symmetry properties of atoms and molecules play an important role in the association or dissociation of molecular complexes. In the framework of the group theory it is possible to predict if the reaction is allowed or not. We did that, using the correlation Tables of \cite{1966msms.book.....H}  and \cite{carter1997molecular}. When symmetry rules avoid the reaction, this one is preceded, in Tables~\ref{ione1} to ~\ref{ioio1}, by the sign $\diamondsuit$.

\end{enumerate}

In Table~\ref{sym} a list of all atoms, molecules and clusters participating to the chemical reactions studied  in the present work together with their electronic ground states and symmetry point groups is given. References are \cite{2005PhRvA..71c3202D}  for C$_n$, \cite{PhDST} for C$_n^+$, \cite{doi:10.1021/jp963200z} for C$_n^-$, \cite{2008JChPh.128l4312T} for C$_n$H et C$n$H$^+$ and the KIDA database for C$_3$H$_2$ and C$_3$H$_2^+$.

\bibliographystyle{apj}

\bibliography{apj2012}

% ------------------- toutes les figures ------------------------------------------------------------------------

\clearpage

%E$_{sat}$ - E$_{dis}$
%(see $/s$ 2.4)
 
% ------------------- toutes les tables  ------------------------------------------------------------------------

\clearpage

\begin{table*}																% Table B1
\caption{Electronic ground states (GS) and symmetry point groups (Sym.) of atoms, molecules and clusters participating in chemical reactions studied in the present work.\label{sym}}             
\centering                          % used for centering table
\begin{tabular}{lcccccccccc}        % centered columns (4 columns)
\hline                 
\textbf{Species}&\textbf{C} &\textbf{C$_2$}& \textbf{C$_3$}& \textbf{C$_4$}& \textbf{C$_5$}& \textbf{C$_6$}& \textbf{C$_7$}& \textbf{C$_8$}& \textbf{C$_9$}& \textbf{C$^+$}\\  
\hline                 

GS & $^3P_g$ & $^1\Sigma_g^+$ & $^1\Sigma_g^+$ & $^1A_g$ & $^1\Sigma_g^+$ & $^1A_1^\prime$ & $^1\Sigma_g^+$ & $^1A_g$ & $^1\Sigma_g^+$ & $^2P_u$ \\

Sym. & atom & $D{_\infty}h$ & $D_{\infty}h$ & $D_2h$ & $D{_\infty}h$ & $D_3h$ & $D{_\infty}h$ & $D_4h$ & $D{_\infty}h$ & atom\\

\hline                 
  
\textbf{Species}&\textbf{C$_2^+$} &\textbf{C$_3^+$}& \textbf{C$_4^+$}& \textbf{C$_5^+$}& \textbf{C$_6^+$}& \textbf{C$_7^+$}& \textbf{C$_8^+$}& \textbf{C$_9^+$}& \textbf{C$^-$}& \textbf{C$_2^-$}\\

GS & $^4\Sigma_g^-$ & $^2B_2$ & $^2A^\prime$ & $^2\Sigma_u^+$ & $^2\Pi_u$ & $^2B_2$ & $^2B_u$ & $^4A_2$ & $^4S_u$& $^2\Sigma_g^+$\\  

Sym. & $D{_\infty}h$ & $C_2v$ & $Cs$ & $D{_\infty}h$ & $D{_\infty}h$ & $C_2v$ & $C_4h$ & $C_2v$ & atom &$D{_\infty}h$ \\
    
\hline  

\textbf{Species}&\textbf{C$_3^-$} &\textbf{C$_4^-$}& \textbf{C$_5^-$}& \textbf{C$_6^-$}& \textbf{C$_7^-$}& \textbf{C$_8^-$}& \textbf{C$_9^-$}& \textbf{CH}& \textbf{C$_2$H}& \textbf{l-C$_3$H}\\

GS & $^2\Pi_g$ & $^2\Pi_g$ & $^2\Pi_u$ & $^2\Pi_u$ & $^2\Pi_u$ & $^2\Pi_g$ & $^2\Pi_u$ & $^2\Pi$ & $^2\Sigma^+$ & $^2\Pi$ \\

Sym. & $D{_\infty}h$ & $D{_\infty}h$ & $D{_\infty}h$ & $D{_\infty}h$ & $D{_\infty}h$ & $D{_\infty}h$ & $D{_\infty}h$ & $C{_\infty}h$ & $C{_\infty}h$ & $C{_\infty}h$ \\

\hline  

\textbf{Species}&\textbf{c-C$_3$H} & \textbf{CH$^+$} & \textbf{C$_2$H$^+$}& \textbf{C$_3$H$^+$}& \textbf{CH$_2$}& \textbf{CH$_2^+$}& \textbf{C$_2$H$_2$}& \textbf{C$_2$H$_2^+$}& \textbf{H}& \textbf{H$_2$}\\

GS & $^2B_2$ & $^1\Sigma^+$ & $^3\Pi$ & $^1\Sigma^+$ & $^3B_1$ & $^2A_1 $ & $^1\Sigma_g^+$ & $^2\Pi_u$ & $^2S_g$ & $^1\Sigma_g^+$\\

Sym. & $C_2v$ & $C{_\infty}v$ & $C{_\infty}v$ & $C{_\infty}v$ & $C_2v$ & $C_2v$ & $C{_\infty}v$ & $C{_\infty}h$ & atom & $D{_\infty}h$ \\

\hline  
                     
\end{tabular}
\end{table*}

\begin{table*}																% Table 1
\caption{Branching ratios (BR) for the dissociative electronic recombination (DR) reactions of C$_n^+$ molecules. BRs predicted by the model are reported in column 2 with in parenthesis part of the estimated errors (see text).  KIDA (on Jan. 2012) , OSU 01-2007 and UdFA06  database BRs are given in column 3, 4 and 5 respectively. Underlined BRs are from dedicated experiments and should be used in chemical models. In column 6, ionization potentials (IP) of the C$_n$ complexes are reported in bold, together with in front of each channel exothermicities. All energies are in eV.\label{DR1}}      
\centering                          % used for centering table
\begin{tabular}{cccccc}        % centered columns (4 columns)
\hline\hline                 % inserts double horizontal lines
Reaction & Model & KIDA & OSU & UdFA & \textbf{IP} / $\Delta$E  \\    % table heading 
\hline\hline                        
   \textbf{C$_4^+$+e$^-\rightarrow$} &  &  &  &  & \textbf{11.5}\\
  C$_3$/C & 0.72($\pm$ 0.02) & \underline{0.6} & \underline{0.6} & \underline{0.6} & 6.48\\
   C$_2$/C$_2$ & 0.28 ($\pm$ 0.02) & \underline{0.4} & \underline{0.4} & \underline{0.4} & 5.41 \\
   
   \textbf{C$_5^+$+e$^-\rightarrow$} &  &  &  &  & \textbf{11.0}\\
   C$_3$/C$_2$  & 0.84($\pm$ 0.02) & 0.85 & 0.5 & 0.5 & 5.19\\
    C$_4$/C  & 0.16($\pm$ 0.02) & 0.15 & 0.5 & 0.5 & 4.03 \\

   \textbf{C$_6^+$+e$^-\rightarrow$} &  &  &  &  & \textbf{9.7}\\
   C$_3$/C$_3 $ & 0.77($\pm$ 0.02) & 0.8 &  &  & 5.70\\
    C$_4$/C$_2 $ & 0.14($\pm$ 0.02) & 0.1 & 0.5 & 0.5 & 3.26\\
    C$_5$/C  & 0.09($\pm$ 0.01) & 0.1 & 0.5 & 0.5 & 4.24 \\
  
 \textbf{C$_7^+$+e$^-\rightarrow$} &  &  &  &  & \textbf{10.1}\\
   C$_4$/C$_3 $ & 0.79($\pm$ 0.01) & 0.8 & 0.14 & 0.14 & 4.59 \\
    C$_5$/C$_2 $ & 0.20($\pm$ 0.01) & 0.2 & 0.43 & 0.43 & 4.30 \\
    C$_6$/C  & 0.01($\pm$ 0.01) &   & 0.43 & 0.43 & 3.77 \\
      
\textbf{C$_8^+$+e$^-\rightarrow$} &  &  &  &  & \textbf{9.2}\\
   C$_5$/C$_3 $ & 0.86($\pm$ 0.03) & 0.8 &  &  & 4.70\\
    C$_4$/C$_4 $ & 0.07($\pm$ 0.02) &   &  &  & 2.90\\
    C$_6$/C$_2 $ & 0.04($\pm$ 0.01) & 0.2 & 0.5 & 0.5 & 2.67\\
   C$_7$/C  & 0.03($\pm$ 0.01) &  & 0.5 & 0.5 & 3.45 \\

  \textbf{C$_9^+$+e$^-\rightarrow$} &  &  &  &  & \textbf{9.4}\\
   C$_6$/C$_3 $ & 0.62($\pm$ 0.02) & 0.65 &  &  & 3.78 \\
    C$_5$/C$_4 $ & 0.31($\pm$ 0.02) & 0.3 &  &  & 3.53 \\
    C$_7$/C$_2 $ & 0.06($\pm$ 0.01) & 0.05 & 0.5 & 0.5 & 3.03\\
    C$_8$/C  & 0.01($\pm$ 0.01) &  & 0.5 & 0.5 & 2.70\\
   
  \textbf{C$_{10}^+$+e$^-\rightarrow$} &  &  &  &  & \textbf{9.2}\\
   C$_7$/C$_3 $ & 0.69($\pm$ 0.02) & 0.75 &  &  & 4.4 \\
    C$_5$/C$_5 $ & 0.27($\pm$ 0.02) & 0.25 &  &  & 4.4\\
    C$_6$/C$_4 $ & 0.04($\pm$ 0.01) &  &  &  & 2.9\\
   C$_9$/C  & 0.0($\pm$ 0.01) &  & 0.5 & 0.5 & 3.2 \\
    C$_8$/C$_2 $ & 0.0($\pm$ 0.01) &  & 0.5 & 0.5 & 2.6 \\ 
\hline                                   
\end{tabular}
\end{table*}

\begin{table*}																% Table 2
\caption{Branching ratios  for the dissociative electronic recombination (DR) reactions of C$_n$H$^+$ molecules. 
Same legends as Table~\ref{DR1}. \label{DR2}}
\centering                          % used for centering table
\begin{tabular}{cccccc}        % centered columns (4 columns)
\hline\hline                 % inserts double horizontal lines
Reaction & Model & KIDA & OSU & UDfA & \textbf{IP} / $\Delta$E  \\    % table heading 
\hline\hline                          
   \textbf{C$_2$H$^+$+e$^-\rightarrow$}  &  &  &  &  & \textbf{11.6}\\
   C$_2$/H  & 0.58($\pm$ 0.05) & $ \underline{0.43} $ & 0.43 & 0.44 & 6.59\\
  C/CH  & 0.34($\pm$ 0.04) & $ \underline{0.39}$ & 0.39 & 0.56 & 3.74\\
   C/C/H  & 0.08($\pm$ 0.05) & $ \underline{0.18} $ & 0.18 &  & 0.1\\
  
   \textbf{C$_3$H$^+$+e$^-\rightarrow$} &  &  &  &  &  \textbf{8.7}\\
   C$_3$/H  & 0.61($\pm$ 0.02) & 0.5 & 0.5 & 0.5 & 5.24\\
    C$_2$H/C   & 0.36($\pm$ 0.02) & 0.5 & 0.5 & 0.5 & 2.49\\
    C$_2$/CH  & 0.03($\pm$ 0.01) &  &  &  & 1.13\\
   
   \textbf{C$_4$H$^+$+e$^-\rightarrow$} &  &  &  &  & \textbf{12.0}\\
   C$_3$/C/H  & 0.40($\pm$ 0.12) &  &  &  & 1.64\\
    C$_2$H/C$_2$  & 0.22($\pm$ 0.06) & 0.3 & 0.3 &  & 5.39\\
    C$_4$/H  & 0.19($\pm$ 0.08) & 0.4 & 0.4 & 1.0 & 7.13\\
    C$_3$H/C  & 0.19($\pm$ 0.09) & 0.15 & 0.15 &  & 5.29\\
    C$_3$/CH  & 0.0($\pm$ 0.001) & 0.15 & 0.15 &  & 6.90\\   
\hline                                   
\end{tabular}
\end{table*}

\begin{table*}																% Table 3
\caption{Branching ratios  for the dissociative electronic recombination (DR) reactions of C$_3$H$_2^+$ molecules. 
Same legend as Table~\ref{DR1}. l referes to linear C$_3$H$_2$ isomers and c to the cyclic one.\label{DR3}} 
%\label{table:3}      % is used to refer this table in the text
\centering                          % used for centering table
\begin{tabular}{cccccc}        % centered columns (4 columns)
\hline\hline                 % inserts double horizontal lines
Reaction & Model & KIDA & OSU & UDfA & \textbf{IP} / $\Delta$E  \\    % table heading 
\hline\hline                        
   \textbf{c-C$_3$H$_2^+$ + $e^- \rightarrow$} & &  &  &  &  \textbf{9.15}\\
    c-C$_3$H/H  & 0.34($\pm$ 0.08) & $ \underline{0.36}$ & 0.36  & 0.14 & 4.78\\
    C$_2$H$_2$/C  & 0.27($\pm$ 0.03) & $ \underline{0.07}$ & 0.07 & 0.14 & 3.22\\
    C$_3$/H$_2$  & 0.18($\pm$ 0.06) & $ \underline{0.36}$ & 0.36 & 0.28 & 5.97\\
    C$_3$/2H  & 0.13($\pm$ 0.03) & $ \underline{0.14}$ & 0.14 & 0.28 & 1.35\\
    C$_2$H/CH  & 0.05($\pm$ 0.01) &  &  &  & 2.41\\
    C$_2$/CH$_2$  & 0.03($\pm$ 0.01) & $ \underline{0.07}$ & 0.07 & 0.14 & 1.55\\

   \textbf{l-C$_3$H$_2^+$ + $e^- \rightarrow$} & &  &  &  &  \textbf{10.4}\\   
    C$_3$/2H  & 0.39($\pm$ 0.04) & 0.14 & 0.14 & 0.28 & 2.60\\
    C$_2$H$_2$/C  & 0.38($\pm$ 0.06) & 0.07 & 0.07 & 0.14 & 4.47\\
    l-C$_3$H/H  & 0.07($\pm$ 0.06) & 0.36 & 0.36 & 0.14 & 6.03\\
    C$_2$H/CH  & 0.07($\pm$ 0.01) &  &  &  & 3.66\\
    C$_2$/CH$_2$  & 0.05($\pm$ 0.01) & 0.07 & 0.07 & 0.14 & 2.80\\
    C$_3$/H$_2$  & 0.04($\pm$ 0.01) & 0.36 & 0.36 & 0.28 & 7.22\\
\hline                                   
\end{tabular}
\end{table*}

\begin{table*}																		   % Table 4
\caption{ Branching ratios  for the charge exchange reactions between C$_n$ and He$^+$.  Same legend as in Table~\ref{DR1} except for the last column. In column 6, \textbf{$\Delta$IP} =  IP(He)-IP(C$_n$) are reported in bold, together with in front of each channel exothermicities. All energies are in eV.\label{He1}}
\centering                          % used for centering table
\begin{tabular}{cccccc}        % centered columns (4 columns)
\hline\hline                 % inserts double horizontal lines
Reaction & Model & KIDA & OSU & UDfA & \textbf{$\Delta$IP} / $\Delta$E  \\    % table heading 
\hline\hline                        
   \textbf{C$_4$ + He$^+ \rightarrow$} & &  &  &  &  \textbf{13.1} \\
    C$_3^+$/C  & 0.61($\pm$ 0.06) & 0.33 & 0.33  & 0.33 & 7.41\\
    C$_3$/C$^+ $  & 0.22($\pm$ 0.03) & 0.33 & 0.33 & 0.33 & 7.87\\
    C$_2$/C$_2^+ $  & 0.16($\pm$ 0.03) & 0.33 & 0.33 & 0.33 & 6.08\\
    C$_2$/C/C$^+ $  & 0.01($\pm$ 0.01) &  &  &  & 0.62\\
                                  
   \textbf{C$_5$ + He$^+ \rightarrow$} & &  &  &  &  \textbf{13.6} \\
    C$_3^+$/C$_2$  & 0.47($\pm$ 0.10) & 0.5 & 0.5  & 0.5 & 7.31\\
    C$_3$/C/C$^+ $  & 0.20($\pm$ 0.10) &  &  &  & 1.77\\
    C$_4^+$/C   & 0.14($\pm$ 0.04) & 0.5 & 0.5 & 0.5 & 7.00\\
    C$_3$/C$_2^+$  & 0.10($\pm$ 0.03) &  &   &  & 7.24\\
    C$_3^+$/C/C   & 0.09($\pm$ 0.06) &  &  &  & 1.31\\
                                  
   \textbf{C$_6$ + He$^+ \rightarrow$} & &  &  &  &  \textbf{14.9} \\
   C$_3^+$/C$_3$  & 0.31($\pm$ 0.10) &  &   &   & 8.68\\
    C$_3^+$/C$_2$/C   & 0.19($\pm$ 0.06) &  &  &  & 1.07\\
    C$_5^+$/C  & 0.16($\pm$ 0.05) & 0.5 & 0.5  & 0.5 & 7.84\\
    C$_3$/C$_2$/C$^+ $  & 0.16($\pm$ 0.04) &  &  &  & 1.55\\
    C$_3$/C$_2^+$/C   & 0.07($\pm$ 0.03) &  &  &  & 1.18\\   
    C$_4^+$/C$_2$  & 0.07($\pm$ 0.04) & 0.5 & 0.5  & 0.5 & 7.18\\
    C$_4^+$/C/C   & 0.04($\pm$ 0.02) &  &  &  & 1.09\\
                                  
   \textbf{C$_7$ + He$^+ \rightarrow$} & &  &  &  &  \textbf{14.5} \\
    C$_4^+$/C$_3$  & 0.40($\pm$ 0.08) &  &   &  & 6.83\\
    C$_5^+$/C$_2 $  & 0.24($\pm$ 0.05) & 0.5 & 0.5 & 0.5 & 5.98\\
    C$_3^+$/C$_3$/C   & 0.14($\pm$ 0.05) &  &  &  & 0.74\\   
    C$_3$/C$_3$/C$^+ $  & 0.10($\pm$ 0.04) &  &  &  & 1.21\\
    C$_4$/C$_3^+$  & 0.08($\pm$ 0.02) &  &   &  & 6.06\\
    C$_6^+$/C  & 0.04($\pm$ 0.01) & 0.5 & 0.5  & 0.5  & 6.96\\
                                  
   \textbf{C$_8$ + He$^+ \rightarrow$} & &  &  &  &  \textbf{15.4} \\
    C$_3^+$/C$_3$/C$_2 $  & 0.47($\pm$ 0.07) &  &  &  & 1.93\\
    C$_4^+$/C$_3$/C   & 0.27($\pm$ 0.05) &  &  &  & 1.95\\
    C$_5^+$/C$_3$  & 0.13($\pm$ 0.09) &  &   &   & 8.31\\   
    C$_5$/C$_2$/C$^+ $  & 0.06($\pm$ 0.03) &  &  &  & 1.17\\
    C$_3$/C$_3$/C$_2^+$  & 0.04($\pm$ 0.01) &  &   &  & 2.03\\
    C$_4$/C$_3$/C$^+ $  & 0.03($\pm$ 0.01) &  &  &  & 1.65\\   
   C$_6^+$/C$_2$  & 0($\pm$ 0.005) & 0.5 & 0.5 & 0.5 & 8.15\\
   C$_7^+$/C   & 0($\pm$ 0.001) & 0.5 & 0.5  & 0.5 & 9.62\\
   
   \textbf{C$_9$ + He$^+ \rightarrow$} & &  &  &  &  \textbf{15.2} \\
    C$_3^+$/C$_3$/C$_3$  & 0.49($\pm$ 0.03) &  &  &  & 2.95\\
    C$_5^+$/C$_3$/C  & 0.29($\pm$ 0.03) &  &  &  & 2.12\\
    C$_4^+$/C$_3$/C$_2$  & 0.20($\pm$ 0.03) &  &  &  & 1.45\\
    C$_6^+$/C$_3$  & 0.02($\pm$ 0.03) &  &  &  & 8.98\\
    C$_8^+$/C  & 0($\pm$ 0.001) & 0.5 & 0.5  & 0.5  & 9.01 \\
    C$_7^+$/C$_2$  & 0($\pm$ 0.001) & 0.5 & 0.5 & 0.5 & 8.92\\

   \textbf{C$_{10}$ + He$^+ \rightarrow$} & &  &  &  &  \textbf{15.4} \\
   C$_7^+$/C$_3$  & 0.34($\pm$ 0.12) & 0.55 &  &  & 8.35\\
   C$_6^+$/C$_3$/C  & 0.23($\pm$ 0.07) & 0.24  &  &  & 2.32\\
   C$_4^+$/C$_3$/C$_3$  & 0.21($\pm$ 0.08) &  &  &  & 0.70\\
   C$_5^+$/C$_5$  & 0.13($\pm$ 0.05) & 0.03 &  &  & 6.15\\
   C$_7^+$/C$_2$/C  & 0.06($\pm$ 0.02) &  &  &  & 1.50\\
  C$_6^+$/C$_4$  & 0.02($\pm$ 0.02) &0.02  &  &  & 6.10\\
   C$_9^+$/C  & 0.01($\pm$ 0.01) & 0.03 & 0.5  & 0.5  & 6.90\\
   C$_8^+$/C$_2$  & 0($\pm$ 0.005) & 0.13 & 0.5 & 0.5 & 8.60\\   
\hline\hline                                   
\end{tabular}
\end{table*}															      	

\begin{table*}	   																	   % Table 5
\caption{ Branching ratios  for the charge exchange reactions between C$_n$H and He$^+$. 
Same legend as Table~\ref{He1}.\label{He2}}
\centering                          % used for centering table
\begin{tabular}{cccccc}        % centered columns (4 columns)
\hline\hline                 % inserts double horizontal lines
Reaction & Model & KIDA & OSU & UDfA & \textbf{$\Delta$IP} / $\Delta$E   \\    % table heading 
\hline\hline                        
   \textbf{C$_2$H + He$^+ \rightarrow$} & &  &  &  &  \textbf{13.0} \\
    C$^+$/C/H  & 0.56($\pm$ 0.04) &  &   &  & 1.84\\
    CH$^+$/C  & 0.18($\pm$ 0.09) & 0.33 & 0.33  & 0.33 & 6.11\\
    C$_2^+$/H  & 0.16($\pm$ 0.08) & 0.33 & 0.33  & 0.33 & 8.19\\
    C$_2$/H$^+$  & 0.07($\pm$ 0.02) &  &   &  & 6.00\\
    C$^+$/CH  & 0.03($\pm$ 0.02) & 0.33 & 0.33  & 0.33 & 5.49\\

   \textbf{c-C$_3$H + He$^+ \rightarrow$} & &  &  &  &  \textbf{15.5} \\
    C$_2^+$/C/H  & 0.45($\pm$ 0.05) &  &   &  & 1.94\\
    C$_2$/C$^+$/H  & 0.29($\pm$ 0.04) &  &   &  & 2.09\\
    C$_3^+$/H  & 0.06($\pm$ 0.04) & 1.0 & 1.0  &  & 9.49\\
   C$_2$H$^+$/C  & 0.06($\pm$ 0.04) &  &   &  & 6.74\\
    C$_2^+$/CH  & 0.05($\pm$ 0.03) &  &   &  & 5.59\\
    C$_2$H/C$^+$  & 0.03($\pm$ 0.02) &  &   &  & 7.09\\
    C$_2$/CH$^+$  & 0.03($\pm$ 0.02) &  &   &  & 6.36\\
    C$_3$/H$^+$  & 0.03($\pm$ 0.01) &  &   &  & 7.50\\
  
   \textbf{l-C$_3$H + He$^+ \rightarrow$} & &  &  &  &  \textbf{16.2} \\
    C$_2^+$/C/H  & 0.54($\pm$ 0.03) &  &   &  & 2.64\\
    C$_2$/C$^+$/H  & 0.35($\pm$ 0.02) &  &   &  & 2.79\\
    C/C/CH$^+$  & 0.07($\pm$ 0.04) &  &   &  & 0.55\\
    C$_3^+$/H  & 0.02($\pm$ 0.02) & 1.0 & 1.0  &  & 10.19\\
    C$_3$/H$^+$  & 0.02($\pm$ 0.01) &  &   &  & 8.20\\

   \textbf{C$_4$H + He$^+ \rightarrow$} & &  &  &  &  \textbf{12.6} \\
    C$_3^+$/C/H  & 0.46($\pm$ 0.05) &  &   &  & 2.63\\
   C$_3$/C$^+$/H  & 0.29($\pm$ 0.03) &  &   &  & 2.98\\
   C$_2^+$/C$_2$/H  & 0.16($\pm$ 0.04) &  &   &  & 1.58\\
   C$_3$H$^+$/C  & 0.04($\pm$ 0.04) &  &   &  & 9.14\\
   C$_2$H$^+$/C$_2$  & 0.04($\pm$ 0.02) & 0.5 & 0.5  & 0.5 & 6.39\\
    C$_4^+$/H  & 0.01($\pm$ 0.02) & 0.5 & 0.5  & 0.5 & 8.78\\
\hline\hline                                   
\end{tabular}
\end{table*}															      	

\begin{table*}	   																	   % Table 6
\caption{ Branching ratios  for the charge exchange reactions between C$_3$H$_2$ and He$^+$. 
Same legend as Table~\ref{He1}.\label{He3}}
\centering                          % used for centering table
\begin{tabular}{cccccc}        % centered columns (4 columns)
\hline\hline                 % inserts double horizontal lines
Reaction & Model & KIDA & OSU & UDfA & \textbf{$\Delta$IP} / $\Delta$E  \\    % table heading 
\hline\hline                        
   \textbf{l-C$_3$H$_2$ + He$^+ \rightarrow$} & &  &  &  &  \textbf{14.15} \\
    C$_2$H$^+$/C/H & 0.25($\pm$ 0.05) &  &   &  & 1.45\\
   C$_3^+$/H/H & 0.24($\pm$ 0.03) &  &   &  & 4.05\\
    C$_2$H/C$^+$/H & 0.15($\pm$ 0.03) &  &   &  & 1.35\\
   C$_3$/H$^+$/H & 0.10($\pm$ 0.02) &  &   &  & 2.05\\
   C$_2$/C$^+$/H$_2$ & 0.10($\pm$ 0.02) &  &   &  & 1.95\\
    C$_2$H$^+$/CH & 0.06($\pm$ 0.02) &  &   &  & 4.95\\
    C$_2$/CH$^+$/H & 0.06($\pm$ 0.03) &  &   &  & 0.75\\
    C$_2$H$_2^+$/C & 0.04($\pm$ 0.01) &  &   &  & 5.95\\
   C$_3$H$^+$/H  & 0.($\pm$ 0.004) & 0.5 & 0.5  & 1.0 & 9.85\\
   C$_3^+$/H$_2$  & 0.($\pm$ 0.002) & 0.5 & 0.5  &  & 8.55\\
   
   \textbf{c-C$_3$H$_2$ + He$^+ \rightarrow$} & &  &  &  &  \textbf{15.43} \\
   C$_2$H$^+$/C/H & 0.28($\pm$ 0.03) &  &   &  & 2.73\\
   C$_2$/CH$^+$/H & 0.16($\pm$ 0.02) &  &   &  & 2.03\\
   C$_3^+$/H/H & 0.14($\pm$ 0.02) &  &   &  & 5.33\\
   C$_2^+$/CH/H & 0.14($\pm$ 0.02) &  &   &  & 1.23\\
   C$_2$H/C$^+$/H & 0.13($\pm$ 0.02) &  &   &  & 3.13\\
    C$_2$/C$^+$/H$_2$ & 0.08($\pm$ 0.01) &  &   &  & 3.23\\
   C$_3$/H$^+$/H & 0.07($\pm$ 0.01) &  &   &  & 3.33\\
   $\rm C_3H^+/H$  & 0.($\pm$ 0.001) & 0.5 & 0.5  & 1.0 & 11.13\\
   $\rm C_3^+/H_2$  & 0.($\pm$ 0.001) & 0.5 & 0.5  &  & 9.83\\
\hline\hline                                   
\end{tabular}
\end{table*}															      
   
\begin{table*}	   																	   % Table 7
\caption{ Branching ratios  for the reactions beetween C$^+$ and C$_n$. Same legend as Table~\ref{DR1}. Pathways indicated by $ \dagger $, $ \S $, or $ \diamondsuit $  are forbidden in ground states (see appendix A). In column 6, the association energies $E_{as}$ (see text),  are reported in bold, together with in front of each channel exothermicities. All energies are in eV.\label{ione1}}      
\centering                          % used for centering table
\begin{tabular}{cccccc}        % centered columns (4 columns)
\hline\hline                 % inserts double horizontal lines
Reaction & Model & KIDA & OSU & UDfA & \textbf{E$_{ass}$} / $\Delta$E  \\    % table heading 
\hline\hline                        
   \textbf{C$_3$ + C $^+ \rightarrow$} & &  &  &  &  \textbf{5.62} \\
   $ \rm C_3/C^+$  & 1.0($\pm$ 0.01) &  &   &  & 0\\
   
   \textbf{C$_4$ + C $^+ \rightarrow$} & &  &  &  &  \textbf{7.04} \\
   $\rm C_3^+/C_2$  & 0.79($\pm$ 0.07) &  &   &  & 0.27 \\
   $ \rm^{\dagger}C_3/C_2^+$  & 0.11($\pm$ 0.05) &  &   &  & 0.38 \\
   $\rm C_4^+/C$    & 0.10($\pm$ 0.05) &  &   &  & 0.29 \\
   $\rm C_5^+$    & 0 & 1.0 & 1.0 & 1.0 & \\ 

   \textbf{C$_5$ + C $^+ \rightarrow$} & &  &  &  &  \textbf{6.97} \\
   $\rm C_3^+/C_3$  & 1.0($\pm$ 0.01) &  &   &  & 0.74 \\
   $\rm C_6^+$    & 0 & 1.0 & 1.0 & 1.0 & \\ 
   
   \textbf{C$_6$ + C $^+ \rightarrow$} & &  &  &  &  \textbf{8.94} \\
   $\rm C_4^+/C_3$  & 0.78($\pm$ 0.04) &  &   &  & 1.27 \\
   $\rm C_5^+/C_2$  & 0.11($\pm$ 0.04) &  &   &  & 0.42 \\
   $\rm C_4/C_3^+$  & 0.06($\pm$ 0.02) &  &   &  & 0.50 \\
   $\rm C_6^+/C$  & 0.05($\pm$ 0.01) &  &   &  & 1.40 \\
   $\rm C_7^+$    & 0 & 1.0 & 1.0 & 1.0 & \\ 
 
   \textbf{C$_7$ + C $^+ \rightarrow$} & &  &  &  &  \textbf{8.05} \\
   $\rm C_5^+/C_3$  & 0.90($\pm$ 0.02) &  &   &  & 1.16 \\
   $\rm C_5/C_3^+$  & 0.05($\pm$ 0.01) &  &   &  & 0.77 \\
   $\rm C_6^+/C_2$  & 0.05($\pm$ 0.01) &  &   &  & 0.62 \\
   $\rm C_8^+$    & 0 & 1.0 & 1.0 & 1.0 & \\ 
   
   \textbf{C$_8$ + C $^+ \rightarrow$} & &  &  &  &  \textbf{8.60} \\
   $\rm C_6^+/C_3$  & 0.66($\pm$ 0.03) &  &   &  & 2.5 \\
   $\rm C_7^+/C_2$  & 0.15($\pm$ 0.01) &  &   &  & 2.3 \\
   $\rm C_5^+/C_4$  & 0.11($\pm$ 0.02) &  &   &  & 1.0 \\
   $\rm C_8^+/C$  & 0.04($\pm$ 0.01) &  &   &  & 2.3 \\
   $\rm C_5/C_4^+$  & 0.04($\pm$ 0.01) &  &   &  & 0.9 \\
   $\rm C_9^+$    & 0 & 1.0 & 1.0 & 1.0 & \\ 
   
   \textbf{C$_9$ + C $^+ \rightarrow$} & &  &  &  &  \textbf{10.4} \\
   $\rm C_{10}^+$    & 0 & 1.0 & 1.0 & 1.0 & \\ 
   $\rm C_7^+/C_3$  & 0.79($\pm$ 0.03) &  &   &  & 3.3 \\
   $\rm C_5^+/C_5$  & 0.12($\pm$ 0.02) &  &   &  & 1.15 \\
   $\rm C_6^+/C_4$  & 0.05($\pm$ 0.02) &  &   &  & 1.1 \\
   $\rm C_9^+/C$  & 0.04($\pm$ 0.02) &  &   &  & 1.9 \\
\hline\hline                                   
\end{tabular}
\end{table*}															     
  
\begin{table*}															% table 8	
\caption{ Branching ratios  for the reactions beetween C$^+$ and C$_n$H. Same legend as Table~\ref{ione1}.\label{ione2}} 
\centering                          % used for centering table
\begin{tabular}{cccccc}        % centered columns (4 columns)
\hline\hline                 % inserts double horizontal lines
Reaction & Model & KIDA & OSU & UDfA & \textbf{E$_{ass}$} / $\Delta$E  \\    % table heading 
\hline\hline                        
   \textbf{CH + C $^+ \rightarrow$} & &  &  &  &  \textbf{7.51} \\
   $\rm CH^+/C$  & 0.53($\pm$ 0.03) & 0.5 & 0.5 & 0.5 & 0.61 \\
   $\rm C_2^+/H$  & 0.43($\pm$ 0.05) & 0.5 & 0.5  & 0.5 & 2.70 \\
   $\rm C_2/H^+$  & 0.04($\pm$ 0.03) &  &   &  & 0.51 \\

   \textbf{C$_2$H + C $^+ \rightarrow$} & &  &  &  &  \textbf{8.41} \\
   $\rm C_3^+/H$  & 0.96($\pm$ 0.02) & 1.0 & 1.0  & 1.0 & 2.40\\
   $\rm C_3/H^+$  & 0.04($\pm$ 0.02) &  &   &  & 0.41\\
   
   \textbf{l{\&}c-C$_3$H + C $^+ \rightarrow$} &  &  &  &  &  \textbf{6.16} \\
   $\rm C_3H^+/C$  & 0.50($\pm$ 0.03) &  &  &  & 2.7\\
   $\rm C_4^+/H$  & 0.35($\pm$ 0.03) & 1.0 & 1.0 &  & 2.34\\
   $\rm C_3/CH^+$  & 0.15($\pm$ 0.03) &  &  &  & 0.81\\

   \textbf{C$_2$H$_2$ + C $^+ \rightarrow$} & &  &  &  &  \textbf{6.7} \\
   $\rm C_3H^+/H$  & 1.0($\pm$ 0.01) & \underline{1.0} & 1.0  & 1.0 & 2.1\\
\hline\hline                                   
\end{tabular}
\end{table*}

\begin{table*}																		   % Table 9   
\caption{ Branching ratios for the reactions beetween H$^+$ and C$_n$ and C$_n$H. Same legend as Table~\ref{ione1}.\label{ione3}} 
\centering                          % used for centering table
\begin{tabular}{cccccc}        % centered columns (4 columns)
\hline\hline                 % inserts double horizontal lines
Reaction & Model & KIDA & OSU & UDfA & \textbf{E$_{ass}$} / $\Delta$E  \\    % table heading 
\hline\hline                        
   \textbf{C$_2$ + H $^+ \rightarrow$} & &  &  &  &  \textbf{7.00} \\
   $\rm ^{\dagger}C_2^+/H$  & 0.80($\pm$ 0.09) & 1.0 & 1.0  & 1.0 & 2.19 \\
   $\rm ^{\dagger}CH^+/C$  & 0.20($\pm$ 0.09) &  &  &   & 0.11 \\

   \textbf{C$_3$ + H $^+ \rightarrow$} & &  &  &  &  \textbf{8.00} \\
   $\rm ^{\S}C_3^+/H$  & 1.00($\pm$ 0.01) & 1.0 & 1.0  & 1.0 & 1.99 \\

   \textbf{C$_4$ + H $^+ \rightarrow$} & &  &  &  &  \textbf{6.47} \\
   $\rm ^{\dagger}C_3H^+/C$  & 0.44($\pm$ 0.04) &  &  &   & 3.01 \\
   $\rm C_4^+/H$  & 0.31($\pm$ 0.03) & 1.0 & 1.0  & 1.0 & 2.65 \\
   $\rm C_3/CH^+$  & 0.15($\pm$ 0.02) &  &  &   & 0.15 \\
   $\rm l{\&}c-C_3H/C^+$  & 0.05($\pm$ 0.05) &  &  &   & 0.31 \\
   
   \textbf{l{\&}c-C$_3$H + H $^+ \rightarrow$} & &  &  &  &  \textbf{8.70} \\
   $\rm ^{\S}C_3H^+/H$  & 0.59($\pm$ 0.04) & 0.5 & 0.5 &   & 4.40 \\
   $\rm C_3^+/H_2$  & 0.28($\pm$ 0.03) & 0.5 & 0.5 &   & 2.50$_*$ \\
   $\rm C_2H_2/C^+$  & 0.05($\pm$ 0.01) &  &  &   & 2.40 \\
   $\rm C_2H_2^+/C$  & 0.04($\pm$ 0.04) &  &  &   & 1.70\\
   $\rm ^{\S\diamondsuit}C_2H/CH^+$  & 0.04($\pm$ 0.04) &  &  &   & 0.47 \\
\hline\hline                                   
\end{tabular}
\end{table*}															      

\begin{table*}	   																	   % Table 10
\caption{Branching ratios for the reactions beetween C$_2^+$ and C$_n$H$_m$. Same legend as Table~\ref{ione1}.\label{ione4}}
\centering                          % used for centering table
\begin{tabular}{cccccc}        % centered columns (4 columns)
\hline\hline                 % inserts double horizontal lines
Reaction & Model & KIDA & OSU & UDfA & \textbf{E$_{ass}$} / $\Delta$E  \\    % table heading 
\hline\hline                        
   \textbf{C + C$_2^+ \rightarrow$} & &  &  &  &  \textbf{7.49} \\
   $\rm ^{\S}C_2/C^+$  & 1.0 & 1.0 & 1.0  & 1.0 & 0.37 \\

   \textbf{C$_2$ + C$_2^+ \rightarrow$} & &  &  &  &  \textbf{7.51} \\
   $\rm C_3^+/C$  & 0.74($\pm$ 0.05) & 1.0 & 1.0  & 1.0 & 1.41 \\
   $ \rm ^{\dagger}C_3/C^+$  & 0.26($\pm$ 0.05) &  &   &  & 1.89 \\

   \textbf{CH + C$_2^+ \rightarrow$} & &  &  &  &  \textbf{9.91} \\
   $\rm C_2H^+/C$  & 0.33($\pm$ 0.03) &  &   &   & 1.15 \\
   $\rm C_3^+/H$  & 0.30($\pm$ 0.02) & 0.5 & 0.5  & 0.5  & 3.90 \\
   $\rm C_2H/C^+$  & 0.19($\pm$ 0.02) &  &   &   & 1.50 \\
   $\rm ^{\dagger}C_2/CH^+$  & 0.15($\pm$ 0.03) & 0.5 & 0.5  & 0.5 & 0.77 \\
   $\rm ^{\dagger}C_3/H^+$  & 0.03($\pm$ 0.01) &  &   &   & 1.91 \\
   
   \textbf{CH$_2$ + C$_2^+ \rightarrow$} & &  &  &  &  \textbf{8.38} \\
   $\rm C_3H^+/H$  & 0.65($\pm$ 0.04) & 0.5 & 0.5 &   & 4.40 \\
   $\rm C_3^+/H_2$  & 0.30($\pm$ 0.04) &  &  &   & 2.50 \\
   $\rm C_2H_2/C^+$  & 0.05($\pm$ 0.01) &  &  &   & 2.40 \\
   $\rm CH_2^+/C_2$  & 0.0($\pm$ 0.001) & 0.5 & 0.5 &   & 0.3 \\
\hline\hline                                   
\end{tabular}
\end{table*}		

\begin{table*}	   																	   % Table 11
\caption{Branching ratios for the reactions beetween CH$^+$ and C$_n$H$_m$. Same legend as Table~\ref{ione1}.\label{ione5}}
\centering                          % used for centering table
\begin{tabular}{cccccc}        % centered columns (4 columns)
\hline\hline                 % inserts double horizontal lines
Reaction & Model & KIDA & OSU & UDfA & \textbf{E$_{ass}$} / $\Delta$E  \\    % table heading 
\hline\hline                        
   \textbf{C + CH$^+ \rightarrow$} & &  &  &  &  \textbf{6.89} \\
   $\rm C_2^+/H$  & 0.95($\pm$ 0.02)  & 1.0 & 1.0  & 1.0 & 2.08 \\
   $\rm ^{\dagger}C_2/H^+$  & 0.05($\pm$ 0.02) &  &   &  & 0.37 \\

   \textbf{C$_2$+CH$^+ \rightarrow$} & &  &  &  &  \textbf{9.14} \\
   $\rm ^{\S}C_3^+/H$  & 0.50($\pm$ 0.05) & 1.0 & 1.0  & 1.0 & 3.13 \\
   $\rm C_2H/C^+$  & 0.27($\pm$ 0.05)  &  &   &  & 0.73 \\
   $\rm C_2H^+/C$  & 0.18($\pm$ 0.07)  &  &   &  &  0.38\\
   $\rm C_3/H^+$  & 0.05($\pm$ 0.02) &  &   &  &     1.14 \\

   \textbf{C$_2$H+CH$^+ \rightarrow$} & &  &  &  &  \textbf{8.2} \\
    $ \rm C_3H^+/H$  & 0.65($\pm$ 0.04) &  &   &  & 3.8 \\
	$\rm ^{\S}C_3^+/H_2$  & 0.30($\pm$ 0.04) & 1.0 &  1.0 & 1.0 & 1.4 \\
	$\rm C_2H_2/C^+$  & 0.05($\pm$ 0.01) &  &   &  & 1.5 \\
\hline\hline                                   
\end{tabular}
\end{table*}		

\begin{table*}	   																	   % Table 12
\caption{ Branching ratios for the  reactions beetween C$_2$H$^+$, C$_2$H$_2^+$ and C or CH. Same legend as Table~\ref{ione1}.\label{ione6}}
\centering                          % used for centering table
\begin{tabular}{cccccc}        % centered columns (4 columns)
\hline\hline                 % inserts double horizontal lines
Reaction & Model & {} & OSU & UDfA & \textbf{E$_{ass}$} / $\Delta$E  \\    % table heading 
\hline\hline                        
   \textbf{C + C$_2$H$^+ \rightarrow$} & &  &  &  &  \textbf{8.76} \\
   $\rm C_3^+/H$  &  0.78($\pm$ 0.07) & 1.0 &  1.0 & 1.0 & 0.37 \\
   $\rm C_2H/C^+$  & 0.16($\pm$ 0.06)  &  &  &  & 0.37 \\
   $\rm C_3/H^+$  & 0.06($\pm$ 0.02) &  &   &  & 2.08 \\

   \textbf{CH + C$_2$H$^+ \rightarrow$} & &  &  &  &  \textbf{9.2} \\
    $\rm C_3H^+/H$  & 0.55($\pm$ 0.04) & 0.5 & 0.5  & 0.5 & 4.9 \\
	$\rm C_3^+/H_2$  & 0.27($\pm$ 0.03) &  &   &  & 2.4 \\
	$\rm C_2H_2^+/C$  & 0.07($\pm$ 0.01) &  &   &  & 0.82 \\
	$\rm C_2H_2/C^+$  & 0.07($\pm$ 0.01) &  &   &  & 1.0 \\
	$\rm C_2H/CH^+$  & 0.04($\pm$ 0.01) &  &   &  & 1.0 \\
	$\rm C_2/CH_2^+$  &  & 0.5 & 0.5  & 0.5 & 1.0 \\
  
   \textbf{C + C$_2$H$_2^+ \rightarrow$} & &  &  &  &  \textbf{8.2} \\
    $\rm C_3H^+/H$  & 0.65($\pm$ 0.04) & 0.33 & 1.0  & 1.0 & 3.8 \\
	$\rm C_3^+/H_2$  & 0.30($\pm$ 0.04) & 0.33 &   &  & 1.4 \\
	$\rm C_2H_2/C^+$  & 0.05($\pm$ 0.01) & 0.33 &   &  & 1.5 \\
\hline\hline                                   
\end{tabular}
\end{table*}		

\begin{table*}	   																	   % Table 13
\caption{Branching ratios (BR) for the neutral-neutral reactions beetween C$_n$H$_m$ and C. Same legend as Table~\ref{DR1}. Pathways indicated by $\dagger$, $\S$, or $\diamondsuit$ are forbidden in ground states (see appendix A). 
In column 6, the association energies (\textbf{$E_{ass}$}) are reported in bold, together with in front of each channel exothermicities. All energies are in eV.\label{nene1}}      
\centering                          % used for centering table
\begin{tabular}{cccccc}        % centered columns (4 columns)
\hline\hline                 % inserts double horizontal lines
Reaction & Model & KIDA & OSU & UDfA & \textbf{E$_{ass}$} / $\Delta$E  \\    % table heading 
\hline\hline                        
   \textbf{C + C$_4 \rightarrow$} & &  &  &  &  \textbf{6.97} \\
   $\rm ^{\dagger}C_3/C_2$  &  1.0 & 1.0 & 1.0 & 1.0 & 1.2 \\

   \textbf{C + C$_5 \rightarrow$} & &  &  &  &  \textbf{5.46} \\
   $\rm ^{\dagger}C_3/C_3$  &  1.0 & 1.0 & 1.0 & 1.0 & 2.5 \\

   \textbf{C + C$_6 \rightarrow$} & &  &  &  &  \textbf{6.33} \\
   $\rm ^{\dagger}C_4/C_3$  &  0.87($\pm$ 0.04) & 0.5 &  0.5 &  & 0.5 \\
   $\rm ^{\dagger}C_5/C_2$  &  0.13($\pm$ 0.04) & 0.5 &  0.5 &  & 0.8  \\

   \textbf{C + C$_7 \rightarrow$} & &  &  &  &  \textbf{5.48} \\
   $\rm ^{\dagger}C_3/C_5$  & 1.0 & 1.0 & 1.0 &  & 1.5 \\

   \textbf{C + C$_8 \rightarrow$} & &  &  &  &  \textbf{6.7} \\
   $\rm ^{\dagger}C_6/C_3$  &  0.67($\pm$ 0.04) & 0.3 & 0.3 &  & 1.1 \\
   $\rm ^{\dagger}C_5/C_4$  &  0.31($\pm$ 0.04) & 0.3 & 0.3 &  &  0.8\\
   $\rm ^{\dagger}C_7/C_2$  &  0.02($\pm$ 0.01) & 0.3 & 0.3 & 1.0 & 0.3 \\

  \textbf{C + C$_9 \rightarrow$} & &  &  &  &  \textbf{6.0} \\
   $\rm ^{\dagger}C_5/C_5$  &  0.87($\pm$ 0.04) &   &   &  &  1.2\\
   $\rm ^{\dagger}C_3/C_7$  &  0.13($\pm$ 0.04) &   &   &  &  1.2\\

   \textbf{C + CH$ \rightarrow$} & &  &  &  &  \textbf{7.86} \\
   $\rm C_2/H $  &  1.0 & 1.0 & 1.0 & 1.0 & 2.8 \\

   \textbf{C + C$_2$H$ \rightarrow$} & &  &  &  &  \textbf{6.21} \\
   $\rm ^{\S}C_3/H $  &  1.0 & 1.0 & 1.0 & 1.0 & 2.75 \\

   \textbf{C + l\&c-C$_3$H$ \rightarrow$} & &  &  &  &  \textbf{6.90} \\
   $\rm C_4/H $  &  0.91($\pm$ 0.04) & 1.0 & 1.0 &  & 2.0 \\
   $\rm C_2H/C_2 $  &  0.09($\pm$ 0.04) &  &  &  & 0.3 \\

   \textbf{C + C$_2$H$_2 \rightarrow$} & &  &  &  &  \textbf{5.93} \\
   $\rm ^{\dagger}C_3/H_2 $  &  0.35($\pm$ 0.04) & \underline{0.73}  & 0.73 & 0.5 & 1.5 \\
   $\rm (l+c)C_3H/H $  &  0.65($\pm$ 0.04) & \underline{0.27}  & 0.27 & 0.5 & 1.5 \\   
\hline\hline                                   
\end{tabular}
\end{table*}		

\begin{table*}	   																	   % Table 14
\caption{Branching ratios for  the neutral-neutal reactions beetween C$_n$H$_m$ and CH or C$_2$H. Same legend as Table~\ref{nene1}.\label{nene2}} 
\centering                          % used for centering table
\begin{tabular}{cccccc}        % centered columns (4 columns)
\hline\hline                 % inserts double horizontal lines
Reaction & model & {} & \textsc{osu} & \textsc{umi} & \textbf{E$_{ass}$}/$\Delta$E  \\    % table heading 
\hline\hline                        
   	\textbf{C$_2$ + CH$ \rightarrow$} & &  &  &  &  \textbf{7.57} \\
   	$\rm C_3/H$  &  0.63($\pm$ 0.04) & 1.0 & 1.0 &  & 4.1 \\
	$\rm C_2H/C$  &  0.37($\pm$ 0.04) &  &  &  & 1.4 \\

   	\textbf{C$_3$ + CH$ \rightarrow$} & &  &  &  &  \textbf{6.71} \\
   	$\rm ^{\S}C_4/H$  &  0.99($\pm$ 0.01) & 1.0 & 1.0 &  & 1.84 \\
   	$\rm C_2H/C_2$  &  0.01($\pm$ 0.01) &  &  &  & 0.1 \\
   	
   	\textbf{C$_2$ + C$_2$H$ \rightarrow$} & &  &  &  &  \textbf{6.61} \\
   	$\rm C_4/H$  &  1.0 & 1.0 & 1.0 &  & 1.74 \\
\hline\hline                                   
\end{tabular}
\end{table*}		

\begin{table*}	   																	   % Table 15
\caption{Branching ratios (BR) for the ion pair reactions beetween C$_n^-$ and C$^+$. 
Same legend as Table~\ref{ione1}.\label{ioio1}}
\centering                          % used for centering table
\begin{tabular}{cccc}        % centered columns (4 columns)
\hline\hline                 % inserts double horizontal lines
Reaction & model & KIDA \& OSU & \textbf{E$_{ass}$}/$\Delta$E  \\    % table heading 
\hline\hline                        
   	\textbf{C$_3^-$ + C$^+ \rightarrow$} & &  &  \textbf{14.33} \\
	$\rm ^{\S}C_2/C/C$  &  0.80($\pm$ 0.10) &  &   2.25 \\
   	$\rm ^{\S}C_3/C$  &  0.15($\pm$ 0.10) & 1.0 &  9.51 \\
	$\rm ^{\S}C_2/C_2$  &  0.05($\pm$ 0.04) &  &   8.24 \\

   	\textbf{C$_4^-$ + C$^+ \rightarrow$} & &  &  \textbf{14.78} \\
  	$\rm ^{\S}C_3/C/C$  &  0.60($\pm$ 0.07) &  &  2.98 \\
  	$\rm ^{\S}C_2/C_2/C$  &  0.34($\pm$ 0.04) &  &  1.72 \\
   	$\rm ^{\S}C_3/C_2$  &  0.06($\pm$ 0.07) &  &  9.27 \\
  	$\rm ^{\S}C_4/C$  &  0.0($\pm$ 0.004) & 1.0 &  7.81 \\

   	\textbf{C$_5^-$ + C$^+ \rightarrow$} & &  &  \textbf{14.18} \\
   	$\rm C_3/C_2/C$  &  0.79($\pm$ 0.05) &  &  2.91 \\ 
   	$\rm C_2/C_2/C_2$  &  0.10($\pm$ 0.03) &  & 1.65  \\
   	$\rm C_4/C/C$  &  0.08($\pm$ 0.03) &  &  1.75 \\
   	$\rm C_3/C_3$  &  0.03($\pm$ 0.05) &  &  10.18 \\
    $\rm C_5/C $  &  0.0($\pm$ 0.007) & 1.0 &  8.72 \\
   		
   	\textbf{C$_6^-$ + C$^+ \rightarrow$} & &  &  \textbf{13.87} \\
   	$\rm C_3/C_3/C$  &  0.53($\pm$ 0.03) &  &  3.54 \\
   	$\rm C_3/C_2/C_2$  &  0.40($\pm$ 0.03) &  &  2.26 \\
   	$\rm C_4/C_2/C$  &  0.05($\pm$ 0.02) &  &  1.10 \\
   	$\rm C_5/C_2$  &  0.02($\pm$ 0.02) &  &  8.07 \\
   	$\rm C_6/C$  &  0.00($\pm$ 0.001) & 1.0 &  7.54 \\

   	\textbf{C$_7^-$ + C$^+ \rightarrow$} & &  &  \textbf{13.99} \\
	$\rm C_3/C_3/C_2$  &  0.72($\pm$ 0.03) &  &  3.11 \\
	$\rm C_4/C_3/C$  &  0.21($\pm$ 0.03) &  &  2.36 \\
	$\rm C_5/C_2/C$  &  0.05($\pm$ 0.02) &  &  1.88 \\
	$\rm C_5/C_3$  &  0.02($\pm$ 0.003) &  &  9.49 \\
   	$\rm C_7/C$  &  0.0($\pm$ 0.003) & 1.0 &  8.24 \\

   	\textbf{C$_8^-$ + C$^+ \rightarrow$} & &  &  \textbf{14.06} \\
	$\rm ^{\S}C_3/C_3/C_3$  &  0.52($\pm$ 0.06) &  &  2.09 \\
	$\rm ^{\S}C_4/C_3/C_2$  &  0.17($\pm$ 0.04) &  &  1.81 \\
	$\rm ^{\S}C_6/C_3$  &  0.15($\pm$ 0.09) &  &  8.44 \\
	$\rm ^{\S}C_5/C_3/C$  &  0.14($\pm$ 0.02) &  &  2.86 \\
	$\rm ^{\S}C_7/C_2$ &  0.02($\pm$ 0.01) &  &  7.69 \\
	$\rm ^{\S}C_8/C$  &  0.0($\pm$ 0.002) & 1.0 &  7.36 \\
	
   	\textbf{C$_9^-$ + C$^+ \rightarrow$} & &  &  \textbf{14.18} \\
	$\rm C_4/C_3/C_3$  &  0.35($\pm$ 0.23) &  &  7.36 \\
	$\rm C_7/C_3$  &  0.30($\pm$ 0.39) &  &  7.36 \\
	$\rm C_5/C_3/C_2$  &  0.20($\pm$ 0.13) &  &  7.36 \\
	$\rm C_5/C_5$  &  0.11($\pm$ 0.04) &  &  7.36 \\
	$\rm C_6/C_3/C$  &  0.02($\pm$ 0.02) &  &  7.36 \\
	$\rm C_5/C_4/C$  &  0.02($\pm$ 0.02) &  &  7.36 \\
	$\rm C_9/C$  &  0.0($\pm$ 0.001) & 1.0 &  7.36 \\
\hline\hline                                   
\end{tabular}
\end{table*}		

\clearpage

\begin{table*}																% Table A1
\caption{Measured branching ratios (BR,\%) for de-excitation channels of C$_4^+$ following
excitation in $\rm C_4^+ - He$ collision ($v=2.6$ velocity atomic units ($au$)). Absolute 1 $\sigma$ errors are given in parenthesis. The dissociation energies  are reported in columns 3 and 6 in eV from \citet{2006BrJPh..36..529D}.\label{C4p}}             
\centering                          % used for centering table
\begin{tabular}{lcclcc}        % centered columns (4 columns)
\hline\hline                 % inserts double horizontal lines
Channel & BR & $E_{diss}$ & Channel & BR & $E_{diss}$ \\    % table heading 
\hline\hline                       
   $\rm C_3^+/C$ & 19.6 (0.8) & 5.7 & $C_3/C^+$ & 7.2 (0.7) & 5.2\\
   $\rm C_2^+/C_2$ & 5.2 (1.2) & 7.0 & $C_2/C^+/C$ & 35.0 (0.7) & 12.5 \\   
   $\rm C_2^+/2C$ & 13.1 (0.4) & 13.0 & $C^+/3C$ & 18.5 (0.8) & 12.8\\   
\hline                                   
\end{tabular}
\end{table*}														

\begin{table*}																% Table A2
\caption{Measured branching ratios (BR,\%) for de-excitation channels of C$_5^+$ following
excitation in $\rm C_5^+ - He$ collision ($v=2.6$  $au$). Same legend as Table~\ref{C4p}.\label{C5p}}             
\centering                          % used for centering table
\begin{tabular}{lcclcc}        % centered columns (4 columns)
\hline\hline                 % inserts double horizontal lines
Channel & BR & $E_{diss}$ & Channel & BR & $E_{diss}$ \\    % table heading 
\hline\hline                       
   $\rm C_3^+/C_2$ & 17.8 (1.1) & 6.8 & $C_3/C_2^+$ & 4.9 (0.4) & 6.7\\
   $\rm C_4^+/C$ & 6.75 (0.4) & 6.7 & $C_4/C^+$ & 1.4 (0.2) & 7.0 \\
     
   $\rm C_3/C^+/C$ & 13.9 (0.8) & 12.4 & $C_3^+/2C$ & 13.2 (0.8) & 12.8\\
   $\rm 2C_2/C^+$ & 5.9 (0.4) & 13.9 & $C_2^+/C_2/C$ & 11.3 (0.5) & 14.3\\
  
   $\rm C_2/C^+/2C$ & 14.9 (0.9) & 20.0 & $C_2^+/3C$ & 5.4 (0.4)  & 20.3\\ 
  
   $\rm C^+/4C$ & 4.9 (0.4) & 26.0    \\ 
\hline                                   
\end{tabular}
\end{table*}				

\begin{table*}																% Table A3
\caption{Measured branching ratios (BR,\%) for de-excitation channels of C$_6^+$ following excitation in $\rm C_6^+ - He$ collision ($v=2.6$  $au$). Same legend as Table~\ref{C4p}.\label{C6p}}             
\centering                          % used for centering table
\begin{tabular}{lcclcc}        % centered columns (4 columns)
\hline\hline                 % inserts double horizontal lines
Channel & BR & $E_{diss}$ & Channel & BR & $E_{diss}$ \\    % table heading 
\hline\hline                       
   $\rm C_3^+/C_3$ & 22.8(2.0) & 6.2 & $C_5^+/C$ & 6.5(0.5) & 7.1\\
   $\rm C_5/C^+$ & 0.4(0.1) & 7.0 & $C_4^+/C_2$ & 4.5(0.5) & 7.7\\
   $\rm C_4/C_2^+$ & 0.7(0.1) & 8.4 & $C_3/C_2/C^+$ & 9.1(0.7) & 13.4\\
   $\rm C_3^+/C_2/C$ & 17.3(1.5) & 13.8 & $C_3/C_2^+/C$ & 5.3(0.5) & 13.7\\
   $\rm C_4^+/2C$ & 4.1(0.4) & 13.8 & $C_4/C^+/C$ & 1.3(0.2) & 14.1\\
   $\rm C_2^+/2C_2$ & 1.9(0.2) & 15.2 & $C_3/C^+/2C$ & 6.8(0.5) & 19.4\\
   $\rm C_3^+/3C$  & 3.9(0.4) & 19.9 & $2C_2/C^+/C$ & 4.9(0.4) & 21.0 \\
   $\rm C_2^+/C_2/2C$ & 4.4(0.4) & 21.3  & $C_2/C^+/3C$ & 4.1(0.4) & 27.0 \\
   $\rm C_2^+/4C$ & 1.3(0.2) & 27.4 & $C^+/5C$ & 1.1(0.2) & 33.1\\
\hline                                   
\end{tabular}
\end{table*}

\begin{table*}																% Table A4
\caption{Measured branching ratios (BR,\%)  for de-excitation channels of C$_7^+$ following
excitation in $\rm C_7^+ - He$ collision ($v=2.6$  $au$).  Same legend as Table~\ref{C4p}.\label{C7p}}             
\centering                          % used for centering table
\begin{tabular}{lcclcc}        % centered columns (4 columns)
\hline\hline                 % inserts double horizontal lines
Channel & BR & $E_{diss}$ & Channel & BR & $E_{diss}$ \\    % table heading 
\hline\hline                       
   $\rm C_6^+/C$ & 1.8(0.2) & 7.5 & $C_4^+/C_3$ & 18.5 (0) & 7.7\\
   $\rm C_4/C_3^+$ & 2.5(0.3) & 8.4 & $C_5^+/C_2$ & 6.7(0.5) & 8.5\\
   $\rm C_5/C_2^+$ & 0.2(0.1) & 8.8 & $C_6^+/C$ & 0.1(0.1) & 8.9\\
   $\rm 2C_3/C^+$ & 5.1(0.4) & 13.3 & $C_3^+/C_3/C$ & 15.5(1.0) & 13.8\\
   $\rm C_5/C^+/C$ & 0.4(0.1) & 14.5 & $C_5^+/2C$ & 3.0(0.3) & 14.6\\
   $\rm C_3/C_2^+/C_2$ & 2.9(0.3) & 15.2 & $C_4^+/C_2/C$ & 4.5(0.4) & 15.3\\
   $\rm C_3^+/2C_2$ & 6.7(0.5) & 15.3 & $C_4/C_2/C^+$ & 1.0(0.1) & 15.6\\
   $\rm C_4/C_2^+/C$ & 0.5(0.1) & 15.9 & $C_3/C_2/C^+/C$ & 7.6(0.5) & 20.9\\
   $\rm C_3/C_2^+/2C $ & 2.5(0.2) & 21.3 & $C_4^+/3C$ & 1.1(0.1) & 21.4\\
   $\rm C_3^+/C_2/2C$ & 7.1(0.3) & 21.4 & $C_4/C^+/2C$ & 0.3(0.1) & 21.6\\
   $\rm 3C_2/C^+$ & 0.8(0.1) & 22.4 & $C_2^+/2C_2/C$ & 1.9(0.2) & 22.8\\
   $\rm C_3/C^+/3C$ & 2.1(0.2) & 27.0 & $C_3^+/4C$ & 1.2(0.1) & 27.5\\
   $\rm 2C_2/C^+/2C$ & 2.7(0.3) & 28.5 & $C_2^+/C_2/3C$ & 1.6(0.2) & 28.8\\
   $\rm C_2/C^+/4C$ & 1.6(0.2) & 34.6 & $C_2^+/5C$ & 0.3(0.1) & 34.9\\
   $\rm C^+/6C$ & 0.3(0.1) & 40.7 \\
\hline                                   
\end{tabular}
\end{table*}

\begin{table*}																% Table A5
\caption{Measured branching ratios (BR,\%) for de-excitation channels of C$_8^+$ following
excitation in $\rm C_8^+ - He$ collision ($v=2.6$  $au$).  Same legend as Table~\ref{C4p}.\label{C8p}}             
\label{table:20}      % is used to refer this table in the text
\centering                          % used for centering table
\begin{tabular}{lcclcc}        % centered columns (4 columns)
\hline\hline                 % inserts double horizontal lines
Channel & BR & $E_{diss}$ & Channel & BR & $E_{diss}$ \\    % table heading 
\hline\hline                       
   $\rm C_7^+/C$ & 2.3(0.3) & 5.8 & $C_5^+/C_3$ & 26.0(1.6) & 6.7\\
   $\rm C_5/C_3^+$ & 1.5(0.1) & 7.1 & $C_6^+/C_2$ & 1.9(0.2) & 7.25\\
   $\rm C_4^+/C_4$ & 1.4(0.1) & 8.1 & $C_6^+/2C$ & 0.7(0.1) & 13.3\\
   $\rm 2C_3/C_2^+$ & 1.6(0.2) & 13.4 & $C_4^+/C_3/C$ & 8.7(0.6) & 13.4\\
   $\rm C_3^+/C_3/C_2$ & 16.2(1.0) & 13.5 & $C_4/C_3/C^+$ & 1.3(0.1) & 13.8\\
   $\rm C_4/C_3^+/C$ & 1.6(0.2) & 14.2 & $C_5/C_2/C^+$ & 0.3(0.1) & 14.2\\
   $\rm  C_5^+/C_2/C$ & 4.4(0.3) & 14.3 & $C_5/C_2^+/C$ & 0.2(0.1) & 14.6\\
   $\rm C_4^+/2C_2$ & 1.1(0.1) & 15.0 & $C_4/C_2^+/C2$ & 0.3(0.1) & 15.6\\
   $\rm 2C_3/C^+/C$ & 3.3(0.3) & 19.1 & $C_3^+/C_3/2C$ & 5.3(0.4) & 19.5\\
   $\rm C_5/C^+/2C$ & 0.1(0.1) & 20.3 & $C_5^+/3C$ & 0.7(0.1) & 20.4\\
   $\rm C_3/2C_2/C^+$ & 1.9(0.2) & 20.6 & $C_3C_2^+/C_2C$ & 2.1(0.2) & 21.0\\
   $\rm C_4^+/C_2/2C$ & 1.7(0.2) & 21.0 & $C_3^+/2C_2/C$ & 4.2(0.3) & 21.1\\
   $\rm C_4C_2/C^+/C$ & 0.5(0.1) & 21.4 & $C_4/C_2^+/2C$ & 0.2(0.1) & 21.7\\
   $\rm C_3C_2/C^+2C$ & 2.9(0.2) & 26.7 & $C_3/C_2^+/3C$ & 0.7(0.1) & 27.0\\
   $\rm C_4^+/4C$ & 0.2(0.1) & 27.1 & $C_3^+/C_2/3C$ & 2.0(0.2) & 27.2\\
   $\rm 3C_2/C^+/C$ & 0.7(0.1) & 28.2 & $C_2^+/2C_22C$ & 0.8(0.1) & 28.6\\
   $\rm C_3/C^+/4C$ & 0.6(0.1) & 32.8 & $C_3^+/5C$ & 0.2(0.1) & 33.2\\
   $\rm 2C_2/C^+/3C$ & 0.9(0.1) & 34.3 & $C_2^+/C2/4C$ & 0.5(0.1) & 34.7\\
   $\rm C_2/C^+/5C$ & 0.5(0.1) & 40.4 & $C^+/7C$ & 0.1(0.1) & 46.4\\   
\hline                                   
\end{tabular}
\end{table*}		

\begin{table*}																% Table A6
\caption{Measured branching ratios (BR,\%) for de-excitation channels of C$_9^+$ following
excitation in $\rm C_9^+ - He$ collision ($v=2.6$  $au$). Same legend as Table~\ref{C4p}.\label{C9p}}             
\centering                          % used for centering table
\begin{tabular}{lcclcc}        % centered columns (4 columns)
\hline\hline                 % inserts double horizontal lines
Channel & BR & $E_{diss}$ & Channel & BR & $E_{diss}$ \\    % table heading 
\hline\hline                       
   $\rm C_8^+/C$ & 0.9(0.1) & 6.2 & $C_7^+/C_2$ & 3.4(0.8) & 6.3\\
   $\rm C_6^+/C_3$ & 17.5(2.8) & 6.2 & $C_5^+/C_4$ & 2.7(0.3) & 7.8\\
   $\rm C_5/C_4^+$ & 1.0(0.1) & 7.7 & $C_7^+/2C$ & 0.8(0.1) & 12.2\\
   $\rm C_6^+/C_2/C$ & 1.3(0.2) & 13.7 & $C_5^+/C_3/C$ & 9.2(0.8) & 13.1\\
   $\rm C_5/C_3^+/C$ & 0.7(0.1) & 13.5 & $C_5/C_3/C^+$ & 0.5(0.1) & 13.2\\
   $\rm C_5^+/2C_2$ & 1.7(0.2) & 14.8 & $C_4^+/C_4/C$ & 0.5(0.1) & 14.5\\
   $\rm C_4^+/C_3/C_2$ & 7.0(0.6) & 14.0 & $C_4/C_3^+/C_2$ & 1.6(0.2) & 14.7\\
   $\rm C_4/C_3/C_2^+$ & 0.4(0.1) & 14.5 & $C_3^+/2C_3$ & 16.4(1.3) & 12.5\\
   $\rm C_6^+/3C$ & 0.3(0.1) & 19.7 & $C_5^+/C_2/2C$ & 1.3(0.1) & 20.7\\   
   $\rm C_5C_2C^+/C$ & 0.2(0.1) & 20.6 & $C_4^+/C_3/2C$ & 2.3(0.2) & 19.8\\
   $\rm C_4/C_3^+/2C$ & 0.6(0.1) & 20.6 & $C_4C_3C^+/C$ & 0.7(0.1) & 20.1\\
   $\rm C_4^+/2C_2/C$ & 0.9(0.1) & 21.3 & $C_4C_2^+C_2/C$ & 0.3(0.1) & 22.3\\
   $\rm C_4/2C_2/C^+$ & 0.2(0.1) & 21.8 & $C_3^+C_3C_2/C$ & 9.2(0.7) & 19.8\\
   $\rm 2C_3/C_2^+/C$ & 1.0(0.2) & 19.8 & $2C_3/C_2/C^+$ & 2.7(0.3) & 19.4\\
   $\rm C_3^+/3C_2$ & 0.8(0.1) & 21.4 & $C_3/C_2^+/2C_2$ & 0.5(0.1) & 21.3\\
   $\rm C_5^+/4C$ & 0.2(0.1) & 26.8 & $C_4^+/C_2/3C$ & 0.5(0.1) & 27.4\\
   $\rm C_4/C_2^+/3C$ & 0.1(0.1) & 28.1 & $C_4C_2C^+/2C$ & 0.3(0.1) & 27.7\\
   $\rm C_3^+/C_3/3C$ & 1.8(0.2) & 25.9 & $2C_3/C^+/2C$ & 1.5(0.2) & 28.5\\
   $\rm C_3^+/2C_2/2C$ & 2.0(0.2) & 27.5 & $C_3C_2^+C2/2C$ & 1.1(0.1) & 27.3\\
   $\rm C_32C_2C^+/C$ & 1.5(0.2) & 27.0 & $C_2^+/3C_2/C$ & 0.2(0.1) & 28.9\\
   $\rm C_4/C^+/4C$ & 0.1(0.1) & 33.8 & $C_3^+/C_2/4C$ & 0.6(0.1) & 32.8\\
   $\rm C_3/C_2^+/4C$ & 0.2(0.1) & 33.4 & $C_3C_2C^+/3C$ & 1.2(0.1) & 33.1\\
   $\rm C_2^+/2C_2/3C$ & 0.3(0.1) & 34.9 & $3C_2/C^+/2C$ & 0.4(0.1) & 34.6\\
   $\rm C_3^+/6C$ & 0.1(0.1) & 39.6 & $C_3/C^+/5C$ & 0.1(0.1) & 39.1\\
   $\rm C_2^+/C_2/5C$ & 0.2(0.1) & 41.0 & $2C_2/C^+/4C$ & 0.5(0.1) & 40.7\\
   $\rm C_2/C^+/6C$ & 0.1(0.1) & 46.7 & $C^+/8C$ & 0(0.1) & 52.8\\
\hline                                   
\end{tabular}
\end{table*}		

\begin{table*}																% Table A7
\caption{Measured branching ratios (BR,\%) for de-excitation channels of C$_{10}^+$ following
excitation in $\rm C_{10}^+ - He$ collision ($v=2.6$  $au$). Same legend as Table~\ref{C4p}.\label{C10p}}             
\centering                          % used for centering table
\begin{tabular}{lcclcc}        % centered columns (4 columns)
\hline\hline                 % inserts double horizontal lines
Channel & BR & $E_{diss}$ & Channel & BR & $E_{diss}$ \\    % table heading 
\hline\hline                       
   $C_9^+/C$ & 0.7(0.2) & 8.5 & $C_8^+/C_2$ & 0.6(0.2) & 6.8\\
   $C_7^+/C_3$ & 19.5(5.3) & 7.1 & $C_7/C_3^+$ & 0.1(0.1) & 9.6\\
   $C_6^+/C_4$ & 1.0(0.2) & 9.3 & $C_5^+/C_5$ & 3.0(0.4) & 9.2\\
   $C_7^+/C_2/C$ & 1.4(0.3) & 14.8 & $C_6^+/C_3/C$ & 5.0(0.6) & 14.6\\
   $C_6^+/2C_2$ & 0.3(0.1) & 14.2 & $C_5^+/C_4/C$ & 0.8(0.2) & 16.3\\
   $C_5/C_4^+/C$ & 0.4(0.1) & 16.2 & $C_5/C_4/C^+$ & 0.1(0.1) & 16.3\\
   $C_5^+/C_3/C_2$ & 10.3(1.1) & 13.7 & $C_5/C_3^+/C_2$ & 1.0(0.2) & 16.0\\
   $C_5/C_3/C_2^+$ & 0.2(0.1) & 15.9 & $C_4^+/C_4/C_2$ & 0.4(0.1) & 15.1\\
   $C_4^+/2C_3$ & 13.6(1.4) & 14.7 & $C_4/C_3^+/C_3$ & 4.1(0.4) & 15.5\\
   $C_7^+/3C$ & 0.3(0.2) & 20.67 & $C_6^+/C_2/2C$ & 0.3(0.1) & 21.44\\
   $C_5^+/C_3/2C$ & 2.2(0.3) & 21.6 & $C_5/C_3^+/2C$ & 0.1(0.1) & 22.5\\
   $C_5C_3C^+/C$ & 0.2(0.1) & 21.6 & $C_5^+/2C_2/C$ & 0.8(0.2) & 21.3\\
   $C_4^+/C_4/2C$ & 0.1(0.1) & 23.0 & $2C_4/C^+/C$ & 0.1(0.1) & 23.3\\
   $C_4^+C_3C_2/C$ & 3.5(0.4) & 20.4 & $C_4C_3^+C_2/C$ & 0.6(0.2) & 23.2\\
   $C_4C_3C_2^+/C$ & 0.3(0.1) & 23.0 & $C_4C_3C_2/C^+$ & 0.5(0.1) & 20.7\\
   $C_4^+/3C_2$ & 0.2(0.1) & 21.9 & $C_3^+/2C_3/C$ & 6.2(0.6) & 20.8\\
   $3C_3/C^+$ & 1.2(0.2) & 20.3 & $C_3^+/C_3/2C_2$ & 4.3(0.5) & 20.4\\
   $2C_3/C_2^+/C_2$ & 0.8(0.2) & 20.4 & $C_5^+/C_2/3C$ & 0.4(0.1) & 29.2\\
   $C_5C_2C^+/2C$ & 0.1(0.1) & 28.4 & $C_4^+/C_3/3C$ & 0.7(0.2) & 28.3\\
   $C_4/C_3^+/3C$ & 0.1(0.1) & 29.1 & $C_4C_3C^+/2C$ & 0.2(0.1) & 28.6\\
   $C_4^+/2C_2/2C$ & 0.5(0.2) & 29.2 & $C_4C_2^+C_2/2C$ & 0.1(0.1) & 29.8\\
   $C_42C_2C^+/C$ & 0.1(0.1) & 28.3 & $C_3^+C_3C_2/2C$ & 3.4(0.4) & 27.7\\
   $2C_3/C_2^+/2C$ & 0.4(0.1) & 28.2 & $2C_3C_2C^+/C$ & 2.1(0) & 26.1\\
   $C_3^+3C_2/C$ & 0.7(0.1) & 28.0 & $C_3C_2^+2C_2/C$ & 0.6(0.2) & 27.9\\
   $C_3/3C_2/C^+$ & 0.4(0.1) & 27.5 & $C_2^+/4C_2$ & 0.1(0.1) & 29.4\\
   $C_5^+/5C$ & 0.1(0.1) & 35.3 & $C_4^+/C_2/4C$ & 0.2(0.1) & 35.9\\
   $C_4C_2C^+/3C$ & 0.1(0.1) & 36.2 & $C_3^+/C_3/4C$ & 0.5(0.1) & 34.4\\
   $2C_3/C^+/3C$ & 0.6(0.1) & 34.0 & $C_3^+/2C_2/3C$ & 0.7(0.1) & 35.9\\
   $C_3C_2^+C_2/3C$ & 0.5(0.1) & 35.8 & $C_32C_2C^+2C$ & 0.7(0.1) & 34.8\\
   $C_2^+/3C_2/2C$ & 0.1(0.1) & 36.7 & $4C_2/C^+/C$ & 0.1(0.1) & 35.2\\
   $C_3^+/C_2/5C$ & 0.3(0.1) & 42.0 & $C_3C_2C^+4C$ & 0.5(0.1) & 41.6\\
   $C_2^+/2C_2/4C$ & 0.2(0.1) & 43.5 & $3C_2/C^+/3C$ & 0.3(0.1) & 43.0\\
   $C_2^+/C_2/6C$ & 0.1(0.1) & 49.5 & $2C_2/C^+/5C$ & 0.1(0.1) & 49.2\\
   $C_2/C^+/7C$ & 0.1(0.1) & 55.2 & $C^+/9C$ & 0 & 61.3\\
\hline                                   
\end{tabular}
\end{table*}		

\begin{table*}																% Table A8
\caption{Measured branching ratios (BR,\%) for de-excitation channels of C$_3$H$_2$ following
charge transfer in $\rm C_3H_2^+$ - He collision ($v=2.6$  $au$). Absolute 1 $\sigma$ errors are given in parenthesis. The energies of dissociation are reported In columns 3 and 6 in eV from \citet{ISI:000074301600017}.\label{C3H2}}             
\centering                          % used for centering tabled
\begin{tabular}{lcclcc}        % centered columns (4 columns)
\hline\hline                 % inserts double horizontal lines
Channel & BR & $E_{diss}$ & Channel & BR & $E_{diss}$ \\    % table heading 
\hline\hline                       
   $C_3H_2$ & 18.8(6.5) &  & $C_3H/H$ & 19.3(5.4) & 4.4\\  
   $C_3/H_2$ & 9.9(3.9) & 4.4 & $C_2H_2/C$ & 9.8 (4.0) & 5.9\\
   $C_2H/CH$ & 1.6(1.6) & 6.7 & $C_2/CH_2$ & 1.4(1.4) & 7.6\\   
   $C_3/2H$ & 15.8(6.2) & 7.8 & $C_2H/C/H$ & 8.2(4.3) & 10.2\\
   $C_2/CH/H$ & 2.1(2.1) & 11.9 & $C_2/C/H_2$ & 0.9(0.9) & 10.1\\   
   $2CH/C$ & 0.1(0.2) & 13.5 & $CH_2/2C$ & 0.1(0.2) & 14.5\\
   $C_2/C/2H$ & 5.2(1.1) & 14.9 & $2C/CH/H$ & 2.8(0.7) & 17.8\\
   $3C/2H$ & 4.0(0.4) & 21.4 &  &  & \\  
\hline                                   
\end{tabular}
\end{table*}		

\begin{table*}																% Table A8
\caption{Measured branching ratios (BR,\%) for de-excitation channels of C$_3$H$_2^+$ following
excitation in $\rm C_3H_2^+$ - He collision ($v=2.6$  $au$). Absolute 1 $\sigma$ errors are given in parenthesis. In columns 3 and 6 are reported the energies of dissociation (eV) using neutral dissociation energies from \citet{ISI:000074301600017} and ionization potentials from \citet{ISI:A1992GY63100014}.\label{C3H2p} }             
\centering                          % used for centering table
\begin{tabular}{lcclcc}        % centered columns (4 columns)
\hline\hline                 % inserts double horizontal lines
Channel & BR & $E_{diss}$ & Channel & BR & $E_{diss}$ \\    % table heading 
\hline\hline                       

  $H/C_3H^+ $ & 19.2(1.0) & 4.3 & $H_2/C_3^+$ & 5.1(0.5) & 5.6\\  
$CH/C_2H^+ $ & 3.3(0.5) & 9.2 & $C/C_2H_2^+$ & 2.3(0.5) & 8.2\\
$C_2H/CH^+ $ & 1.9(0.4) & 8.2 & $C_2H_2/C^+$ & 1.6(0.3) & 8.0\\
 $ CH_2/C_2^+$ & 0.4(0.2) & 9.9 & $C_2/CH_2^+$ & 0.4(0.1) & 8.8\\
 $C_3H/H^+$ & 0.1(0.2) & 8.7 & $2H/C_3^+$ & 6.8(0.5) & 10.1\\
 $C/H/C_2H^+$ & 6.3(0.6) & 12.7 & $C/CH/CH^+$ & 5.2(1.0) & 16.1\\
 $CH/H/C_2^+$ & 4.1(0.4) & 14.2 & $C_2/H/CH^+$ & 3.4(0.4) & 13.4\\
 $C_2H/H/C^+ $ & 3.2(0.7) & 12.3 & $C_2H/C/H^+$ & 3.2(0.4) & 14.6\\
 $C_3/H/H^+ $ & 2.6(0.3) & 12.1 & $2CH/C^+$ & 2.6(0.5) & 16.7\\
 $C_2/H_2/C^+ $ & 2.5(0.2) & 12.2 & $C/CH_2/C^+$ & 1.3(0.4) & 15.6\\
$C/H_2/C_2^+ $ & 0.8(0.1) & 12.5 & $C_2/CH/H+$ & 0.6(0.2) & 16.4\\
$2C/CH_2^+ $ & 0.3(0.3) & 14.6 & $C/2H/C_2^+$ & 6.1(0.3) & 16.8\\
$C_2/2H/C^+ $ & 4.1(0.1) & 16.7 & $C_2/C/H/H^+$ & 2.5(0.4) & 19.0\\
$CH/C/H/C^+ $ & 2.2(0.7) & 19.8 & $2C/H/CH^+$ & 2.1(0.5) & 19.2\\
$2C/H_2/C^+ $ & 1.4(0.3) & 17.9 & $2C/CH/H^+$ & 0.3(0.4) & 22.2\\
$2C/2H/C^+ $ & 2.6(0.3) & 24.3 & $3C/H/H^+$ & 1.7(0.2) & 26.7\\

\hline                                   
\end{tabular}
\end{table*}

\end{document}